\documentclass[a4paper, 10pt, conference]{ieeeconf}      

\IEEEoverridecommandlockouts                              
\usepackage{graphicx} 
\usepackage{amsmath} 
\usepackage{amssymb}  

\title{\LARGE \bf
Scalar Field Wave (Fuzzy) Dark Matter, the Baryonic Tully-Fisher Relation, and the Formation of Galaxies
}
\author{Benjamin Hamm$^*$ \thanks{$^*$email: bih@duke.edu}\\
\textit{Duke University Department of Physics}}
\begin{document}

\maketitle
\thispagestyle{plain}
\pagestyle{plain}

\begin{abstract}

We investigate Scalar Field Wave Dark Matter in the context of galactic Dark Matter halos.  In particular, we offer an analysis of the Baryonic Tully-Fisher Relation (BTFR).  We detail a particular family of excited state solutions to the Einstein-Klein-Gordon equations, and use it to provide a novel theoretical model for producing the BTFR.  We then solve this model computationally to simulate the BTFR. Interpreting the Dark Matter scalar field as an ultralight boson, this returns a conservative mass constraint of $m\geq 10^{-23}eV$.  Assuming slightly stronger conditions suggests $m\geq 10^{-22}eV$ to be more compatible with the BTFR.  We provide a discussion of Scalar Field Dark Matter rotation curves and the structure of Scalar Field Dark Matter halos.  Compatibility with the BTFR requires the excited state solutions to obey particular boundary conditions; this may have implications for the behavior of Dark Matter halos and the formation of galaxies.     

\end{abstract}

\section{Introduction and Background}
\label{Sec:Intro}
\subsection{Wave Dark Matter ($\psi$DM)}
\label{Sec:WDM}
$\Lambda$ Cold Dark Matter ($\Lambda$CDM) is the most widely studied theory in modern cosmology.  To date, simulations such as IllustrisTNG \cite{IllustrisTNGRelease} and EAGLE \cite{EAGLERelease} have used the $\Lambda$CDM paradigm to model many of the largest scale features of our universe such as the galactic filaments and the cosmic microwave background.  In short, $\Lambda$CDM is the hypothesis that the energy content of the universe consists of about 68\% Dark Energy (represented by the cosmological constant $\Lambda$), 28\% Dark Matter (DM), and 4\% Baryonic Matter.  The existence of DM, matter which interacts via gravitation but not via electromagnetism, was suggested by Fritz Zwicky in 1937 to explain the unexpectedly large velocities of the galaxies within Coma Cluster \cite{Zwicky37}.  Now, DM is an essential feature of cosmological theories, and understanding its properties is a crucial aspect of galactic dynamics and large scale cosmology.  Moreover, alternate theories to the existence of DM, such as Modified Newtonian Dynamics (MoND), now face strong tension with observations, one of the most prominent being that of the Bullet Cluster \cite{BulletCluster1}\cite{BulletCluster2}.  

$\Lambda$CDM as described places relatively few constraints on the composition of DM.  In fact, the name $\Lambda$CDM merely suggests that DM moves at non-relativistic speeds, or in other words that it is \textit{cold} DM (CDM).  A candidate for CDM which has received large amounts of attention is the Weakly Interacting Massive Particle (WIMP), a particle in the mass range of 10-1000GeV which interacts via the Weak Nuclear Force but not the Electromagnetic Force.  Though successful at describing structure at the largest cosmological scales, the WIMP theory displays some shortcomings at smaller scales closer to those of galaxies.   One potential source of these shortcomings is the WIMP matter power spectrum.  This spectrum can extend with significant power to scales smaller than stars \cite{WIMPPower}. WIMP based simulations thus display structure formation on these small scales.  This causes issues in that such small scale structures have not been detected or observed in reality.  The ``missing satellites problem", ``cusp-core problem", and ``too-big-to-fail problem" are a few such issues often tied to this discrepancy.

According to  \cite{NoMissingSatellites} and \cite{TooBigToFail}, issues like the ``missing satellites problem" and the ``too-big-to-fail problem," can be corrected by careful consideration of detection efficiency, and satellite statistics.  Such analysis depends on the chosen models of galaxy-halo connection as well as other disruptive processes like baryon feedback.  A greater understanding of DM physics at the galaxy halo scale would greatly compliment these resolutions, and perhaps allow one to place further constraints on the nature of DM and its interactions with baryons.  

Attempts to resolve these small scale issues often focus on improving DM simulations by introducing mechanisms meant to suppress the power in the small mass end of the DM power spectrum.  For instance, Hot Dark Matter (HDM) theories have been used to achieve this suppression. The high velocity of HDM results in a large free-streaming length for DM particles; this prevents structure formation on small scales.  However, HDM suppresses small scale structure formation so well that it even prevents formation on the scale of galaxies.  For this reason HDM is usually ruled out as a DM candidate \cite{HotDMReview}.  Warm Dark Matter (WDM) is usually ruled out for the same reason.  However, this assumes all DM to be the same form of matter.  Mixtures of different HDM and CDM components may be able to alleviate some of the issues faced by HDM and WDM \cite{HotDMReview}.  Moreover, alternatives to the WIMP paradigm are also considered since searches for WIMP-like particles, to date, have yielded no strong evidence for their existence.

A notable alternate to the WIMP is an ultralight scalar particle with a mass parameter of $m\sim10^{-22}$eV\cite{WittenReview}.   An interesting feature of these ultralight particles is their large deBroglie wavelength scale ($\lambda\sim1$kpc)  which arises as a result of their small mass scale.  It was originally noted in \cite{HuOriginal} that employing an ultralight scalar particle as DM could introduce a sharp cutoff in the low mass end of the power spectrum.  This is due to the repulsive pressure of the scalar field which occurs at small length and high density scales; such pressure inhibits structure formation at the smallest length scales.  Furthermore, these particles are expected to exhibit coherent wavelike features on the galactic scale, which could potentially resolve issues such as the ``cusp-core problem"\cite{WittenReview}.  In particular, dark matter halos formed by ultralight scalars display solitonic, finite density cores as opposed to cuspy ones.  These small scale features and differences provide hope for constraining the properties of such an ultralight scalar by comparing to observational data from the galactic scale. 

Theories regarding these ultralight scalars have been given a myriad of names: Fuzzy Dark Matter (FDM), Wave Dark Matter($\psi$DM), Axion Dark Matter (ADM), Bose-Einstein Condensate Dark Matter (BECDM) and Scalar Field Dark Matter (SFDM) being a few common ones.  FDM, coined by Hu\cite{HuOriginal}, highlights the ultradispersed wavelength of the particle, hence ``fuzzy." ADM pays tribute to the QCD Axion which could potentially resolve the strong CP problem \cite{CPWeinberg}. BECDM highlights the feature that the particles are expected to form large scale condensate structures with superfluid properties \cite{ReviewSuarezRoblesMatos,BECDMCosmology}. Lastly, SFDM hightlights that these particles are described by a scalar field wave equation.

In these cases, dark matter particles are usually described by the coupled Einstein Klein-Gordon equations (EKGEs) and their non-relativistic analog, the Poisson-Schrodinger equations (PSEs); it is their galactic scale, wavelike features, which give these theories their unique character.  In homage to these wavelike features, we choose to use the name Wave ($\psi$) Dark Matter ($\psi$DM).  The name $\psi$DM in itself does not require DM to be an ultralight scalar; one could suppose DM to consist of some other sort of ultralight matter field, a vector for instance.  We focus only on the scalar case.  We do note that in the most specific sense, this paper focuses on the theory of \textit{Scalar Field Wave Dark Matter} (SF$\psi$DM).  In particular we will discuss SF$\psi$DM with a mass parameter of $m\sim10^{-22}eV$.    

In cosmological simulations which include SF$\psi$DM, DM forms a large condensed wave structure in the early universe, usually referred to as a superfluid or Bose-Einstein Condensate \cite{SchiveCosmology}. Fluctuations in the DM density can eventually lead to regions of gravitational collapse and ultimately a condensation like process\cite{Condensation}; as collapse ensues, the outwards pressure of the scalar wave will increase and eventually reach an equilibrium with the inwards gravity, resulting in a stable droplet structure referred to as a $\psi$DM soliton or ground state boson star.  As the universe expands and cools, this soliton condensation will occur; this phenomena is indeed demonstrated in simulations \cite{MoczSolitons}.  Further, on large scales, these DM solitons evolve due to gravity in a particle like fashion, much like DM in WIMP $\Lambda$CDM \cite{SchiveCosmology}.  Thus, in adopting $\psi$DM, one retains the prior features of $\Lambda$CDM in modelling large scale structure, but gains a complicated wave dynamic which must be resolved on the scale of galaxies.  

The formation of stable boson stars in SF$\psi$DM suggests the possibility that galactic DM halos are themselves a type of dynamical boson star system \cite{JaeWeonBosonStar}.  One possibility that has been investigated with high resolution simulation is that galactic scale DM halos may form in a bottom up fashion from the merging of $\psi$DM solitons\cite{GalacticCollapse,MoczGalaxy}.  This gives the resulting galactic halo a rich structure, resulting from the overlap and interference of its many parent solitons.  Generically, the central regions of the halo will display a steep, high, but finite density core which is then surrounded by a turbulent and granular halo.  In the far-from-center and turbulent regions of the halo, the profile converges to be consistent with N-body gravity, agreeing with an Navarro-Frenk-White (NFW) like profile \cite{MoczGalaxy}\cite{SchiveCosmology}.  

In this paper, we will consider DM halos with the interpretation that they are formed in such a bottom up fashion from many parent solitons.  Specifically, we will explore \textit{approximating} the inner region of the halos with spherically symmetric and static (SSS) SF$\psi$DM excited states. SSS SF$\psi$DM excited states, discussed in depth in the following sections, are the mathematical representations of boson stars, and are interesting in many regards.  Particularly, such excited states display an inner core region, which is then surrounded by fluctuations on length scales similar to the core size.  Moreover, SF$\psi$DM excited states display approximately flat rotation curves which have properties compatible with the Baryonic-Tully-Fisher-Relation (BTFR), at least in a DM-only setting\cite{GoetzThesis}.  This feature, detailed in section \ref{Sec:TFRandWDM}, provides a method of describing the BTFR which is new and unique to $\psi$DM.  We will investigate this feature in more depth, and utilize the BTFR to place constraints on the mass parameter of SF$\psi$DM. We will then further discuss the validity of describing galactic halos as SF$\psi$DM excited states.

\subsection{$\psi$DM and the Einstein-Klein-Gordon Equations}
\label{WDMandEKGE}

We now present the mathematical model of SF$\psi$DM.  The governing equations of SF$\psi$DM are the Einstein-Klein-Gordon Equations (EKGEs) for a scalar field.  The EKGEs can be attained for either a complex or real scalar field, $\psi$. There are thus two possible interpretations of  SF$\psi$DM.  In both cases, the low-field, non-relativistic limit is the same.  This can be understood by considering the EKG action for a scalar field \footnote{In the remainder of the paper, equations will be posed in geometric units of $c=G=\hbar=1$.}, $\psi$:
\begin{equation}
\label{eq:EKGAction}
S = \int{\left(R-2\Lambda -16\pi\left(\frac{|d\psi|^2}{2m^2}+\frac{|\psi|^2}{2}\right)\right)dV}
\end{equation}
Here, $R$ is the Ricci Scalar Curvature, $\Lambda$ is the cosmological constant, and $m$ is the mass parameter for $\psi$.  The critical point of this action is the set of EKGEs
\begin{equation}
\label{eq:EinsteinEq}
G + \Lambda g = 8\pi\left(\frac{d\psi\otimes d\bar{\psi} + d\bar{\psi}\otimes d\psi}{2m^2} - \left(\frac{|d\psi|^2}{2m^2}+\frac{|\psi|^2}{2}\right)g\right)
\end{equation}
\begin{equation}
\label{eq:KleinGordonEq}
\Box \psi = m^2\psi
\end{equation}
where $G$ is the Einstein curvature tensor, $g$ is the metric tensor, and $\Box$ is the d'Alemberian wave operator \footnote{We denote the modulus or absolute value of a variable $x$ as $|x|$ and its complex conjugate as $\bar{x}$.  $\otimes$ denotes a tensor product.}.  In the galactic regime, it is a great simplification to approximate with $\Lambda<<1$.  This effectively claims the dark energy contribution to be negligible on the galactic scale.  In other words, on small local scales the expansion rate of spacetime is taken to be negligible in comparison to the dominant local gravity.  Therefore, in the remainder of our discussion, we will neglect the effects of $\Lambda$.

If one interprets $\psi$ as an axion-like field, $\psi$ is necessarily a real scalar \cite{BraatenAxionStar} at the action level.  In this case, equation (\ref{eq:EKGAction}) represents an effective action for the axion field $\psi$.  The non-relativistic effective limit for the real axion field can be expressed in terms of a complex scalar field $\phi$:
\begin{equation}
\label{axionNREFT}
\psi(\vec{r},t) = \frac{1}{\sqrt{2}}\left(\phi(\vec{r},t)e^{imt}+\bar{\phi}(\vec{r},t)e^{-imt}\right)
\end{equation}
Combining this form with the EKGEs, the condition that $E_{\phi}=i\frac{\partial \phi}{\partial t}<<m$ and the low-field representation of the metric line element, $ds^2$, results in the Poisson-Schrodinger (PS) system for a complex scalar.  That is, given $V(r)<<1,$ and
\begin{equation}
\label{eq:LowFieldMetric}
ds^2 = -(1+2V(\vec{r}))dt^2 + (1-2V(\vec{r}))(dx^2 + dy^2 + dz^2)
\end{equation}
we obtain the PS system as follows:
\begin{equation}
i\frac{\partial\phi}{\partial t} =- \frac{1}{2m}\nabla^2\phi + mV\phi
\label{eq:Schrodinger}
\end{equation}
\begin{equation}
\label{eq:Poisson}
\nabla^2V = 4\pi |\phi|^2
\end{equation}
This system represents the low-field, non-relativisic limit of an axion-like particle described by a complex scalar field $\psi$, effectively represented by a real scalar $\phi$.  

There is another way to attain the PS system, which does not interpret $\psi$ as an axion-like real scalar, but instead as a complex scalar.  In the case that we interpret $\psi$ to be a complex scalar, we still attain equations (\ref{eq:KleinGordonEq}) and (\ref{eq:EinsteinEq}) as the critical points of the action.  Then, we restrict to the static low field metric, (\ref{eq:LowFieldMetric}), and take an ansatz similar to (\ref{axionNREFT}), namely:
\begin{equation}
\psi(\vec{r},t) = \phi(\vec{r},t)e^{imt}
\end{equation}
This, combined with the small energy condition, results in the PS form found in equations (\ref{eq:Schrodinger}) and (\ref{eq:Poisson}).  As demonstrated then, the interpretations of SF$\psi$DM as a real scalar or as a complex scalar coincides in the low-field non-relativistic regime, and can be effectively represented by a complex scalar field $\phi$.  To further generalize this low-field limit, one may remove the restriction that $V$ be static.  This introduces terms of $\frac{\partial V}{\partial t}$ in the Schrodinger equation; such a generalized set of equations is referred to as the Gross-Pitaveskii (GP) equations.  Qualitatively the GP equations coupled to the Einstein equations then describe a dynamical self gravitating scalar wave.  

\section{$\psi$DM in Spherical Symmetry}
\label{Sec:WDMinSS}
\subsection{SSS $\psi$DM Excited States}
\label{Sec:SSSExcitedStates}
In this paper, we will consider solutions to the EKGEs which are spherically symmetric and static (SSS).  Moreover, for simplicity, we restrict to the low-field PS case, and take the following ansatz for a \textit{complex} scalar, $\phi$ (note that this case will coincide with the case of a real scalar in the low-field, non-relativistic limit).  
\begin{equation}
\phi(\vec{r},t) = \Phi(r)e^{i(\omega-m)t}
\end{equation}
We note that this is equivalent to considering a SSS harmonic ansatz for the complex case of the low-field, non-relativistic EKGEs, where $\psi(r,t)=\Phi(r)e^{i\omega t}$.
Combining this with equations (\ref{eq:Schrodinger}) and (\ref{eq:Poisson}) results in the following set of ODEs representing the PS system.  
\begin{equation}
\label{eq:SSSSchrodinger}
\frac{1}{2m}\left(\frac{d^2\Phi}{d r^2}+\frac{2}{r}\frac{d\Phi}{d r}\right) = (m-\omega+mV)\Phi
\end{equation}
\begin{equation}
\label{eq:SSSPotential}
\frac{d V}{d r} = \frac{M(r)}{r}
\end{equation}
\begin{equation}
\label{eq:SSSMass}
\frac{d M}{d r} = 4\pi r^2\Phi^2
\end{equation}
Here, $V(r)$ corresponds to what is usually interpreted as a gravitational potential, $\omega$ is a real frequency, and $M(r)$ corresponds to the mass enclosed by a sphere of radius $r$.  This SSS description is commonly used in studies of self-bound boson systems referred to as boson stars.  Boson stars are indeed low-field descriptions of the real valued EKGEs, as well as the complex case.      

\begin{figure}
    \centering
    \includegraphics[height=3in,width=3.35in]{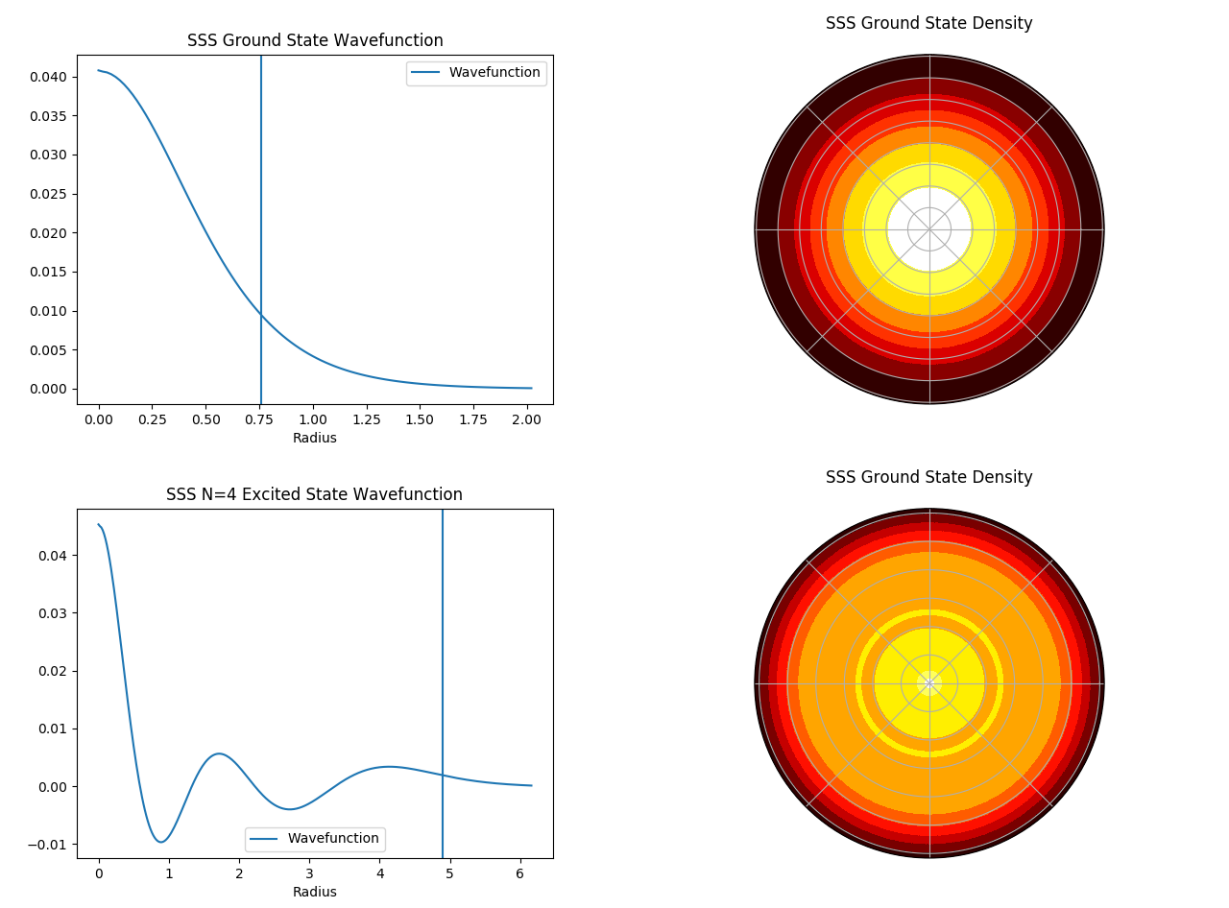}
    \caption{Plots generated with units of $\hbar=c=G=1$, $m=100$ and frequency of $\omega=99.9$.  Top left:   Radial wave function for SSS ground state.  Top right:  Density function for the ground state projected along the line of sight.  Bottom left: N=4 Excited SSS state radial wave function.  Bottom right: N=4 Excited state density projected along the line of sight.  Vertical lines denote the decay radius $R_d$ for each state. Color scale chosen to optimize contrast. }
    \label{fig:SSSExample}
\end{figure}

For a fixed and finite value of total mass, $M_{tot} = \lim_{r\to \infty}M(r)$, there are a countable number of solutions to the the SSS PSEs (specifically, they are countable up to adding a constant to the potential $V(r)$).  Further, these solutions can be categorized by the number of nodes, $n$, which occur before their wavefunction's decay radius $R_d$. After $n$ nodes, the wavefunction will exhibit an exponential decay towards zero value.  Examples of such states are plotted in figure \ref{fig:SSSExample}.  The $n=0$ solution, referred to as the \textit{ground state} solution or the $\psi$DM soliton, has received sizeable attention in studies of SF$\psi$DM.  This is due to the fact that it is a stable attractor solution to the PSEs;  in fact, in the PS limit, the ground state solution has been proven analytically to be stable \cite{MarzuolaExternal}.  Furthermore, simulations show that other, unstable states will tend to evolve towards becoming a ground state after a sufficiently long time \cite{BosonStarEvolution}.  One peculiar property of the the PSEs is that their solutions admit several scaling relations; these are detailed in a later section.  One particular result of the scaling relations is that the mass scale of a particular ground state completely determines its corresponding length scale.  Citing \cite{WittenReview}, the radius containing half of the ground state mass, $R_{1/2}$, is related to the total mass $M$, as
\begin{equation}
\label{eq:GSMass}
    R_{1/2} \approx 0.335 kpc \frac{10^9M_{\odot}}{M}\left(\frac{10^{-22}eV}{m}\right)^2
\end{equation}

 The remaining solutions with $n\geq1$ are referred to as \textit{excited state} solutions of order $n$. Numerical simulations have revealed the SSS excited states to be unstable in time upon applying small perturbations, at least in a DM-only setting.  Moreover, analytical studies of the PSEs have yielded proof that all DM-only states with $n\geq1$ are themselves unstable \cite{MarzuolaExternal}.  On the other hand, including an external gravitational potential can actually stabilize certain excited states\cite{MarzuolaExternal}.  This indicates that the instability of the excited states can be remedied or at least reduced by accounting for the presence of other potentials, or matter, in the background.  Qualitatively, other matter in the background may be able to generate additional gravity to hold the DM excited states in place, or at least allow it to be gravitationally bound.  
 
 To further address the concern of instability, we note that in the following models of galactic dark matter halos, we in no way enforce that galaxies exist in stable and static configurations.  It would seem extremely unlikely that a complex dynamical system such as a galaxy would lie in a perfectly spherically symmetric, stable, and static configuration.  The presence of dynamical processes such as baryon feedback or the in-fall of other matter would likely disrupt the halo and perhaps cause excitation.  We merely suggest galactic halos may be \textit{approximated} to leading order by SSS excited states; this model is detailed in the following sections.

\begin{figure}
    \centering
    \includegraphics[height=1.5in,width=3in]{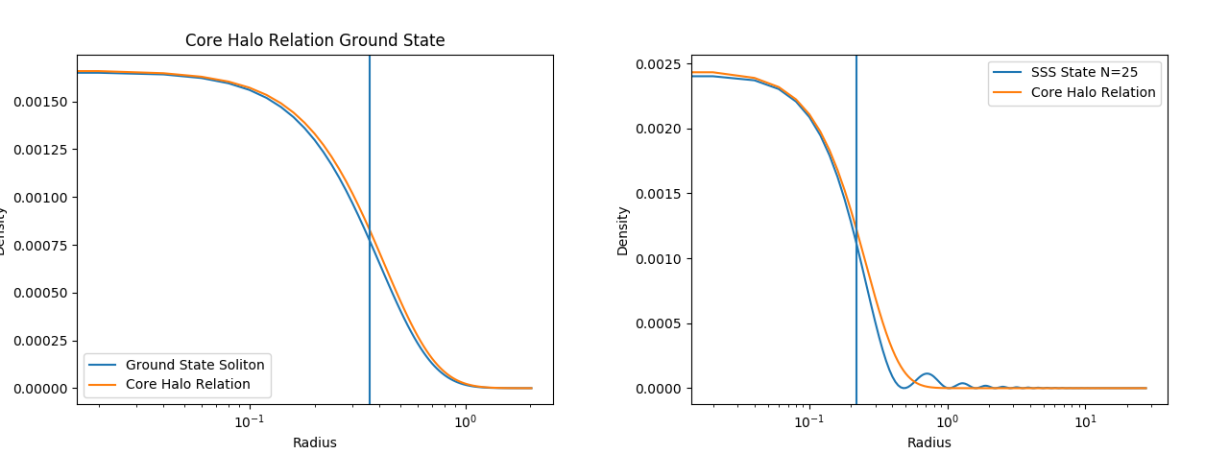}
    \caption{Excited State Core Halo Relation.  Plots rendered using static state frequencies of $\omega=.999 m$, and units of $m=100$.  Left: The core halo relation, (\ref{eq:CoreHalo}), plotted with the ground state density profile.  Right: The core halo relation for an excited state $N=25$, also generated with equation (\ref{eq:CoreHalo}).  Vertical lines denote the value of $r_c$.  Both states display a central, high density core region before $r_c$, and quickly drop off thereafter.  Excited state displays greater spatial extent due to its oscillating region.}
    \label{fig:CoreHaloRelation}
\end{figure}

 We further note that, as seen and described in figure (\ref{fig:CoreHaloRelation}), SSS excited states provide excellent agreement with the core-halo relation for SF$\psi$DM galactic halos (described in \cite{CoreHaloRelation}),at least to leading order.  
 \begin{equation}
 \label{eq:CoreHalo}
     \rho(r) = \rho_0\left(1+.091\left(\frac{r}{r_c}\right)^2\right)^{-8}
 \end{equation}
 Here, $r_c$ denotes the radius at which the density of the DM halo reaches half of its central value.  Apparently the core profiles of excited states are close matches to those of ground state solitons.  As seen in figure (\ref{fig:CoreHaloRelation}), the central region of an excited state closely resembles a soliton core.  The primary difference from the ground state is in the far regions of the halo; excited states display a generally more extended halo.  It thus seems plausible that SF$\psi$DM galaxies could be approximated, at least in the central regions, by SSS excited states.     
 
\subsection{SSS SF$\psi$DM halos}
\label{Sec:SSSHalos}

In this section, we discuss physical properties of the SSS solutions to the PSEs in the context of dark matter halos.  Firstly, we take the SSS form of the PSEs displayed in equations (\ref{eq:SSSSchrodinger}),(\ref{eq:SSSPotential}) and (\ref{eq:SSSMass}).  

\begin{figure}
    \centering
    \includegraphics[height=3in,width=3.35in]{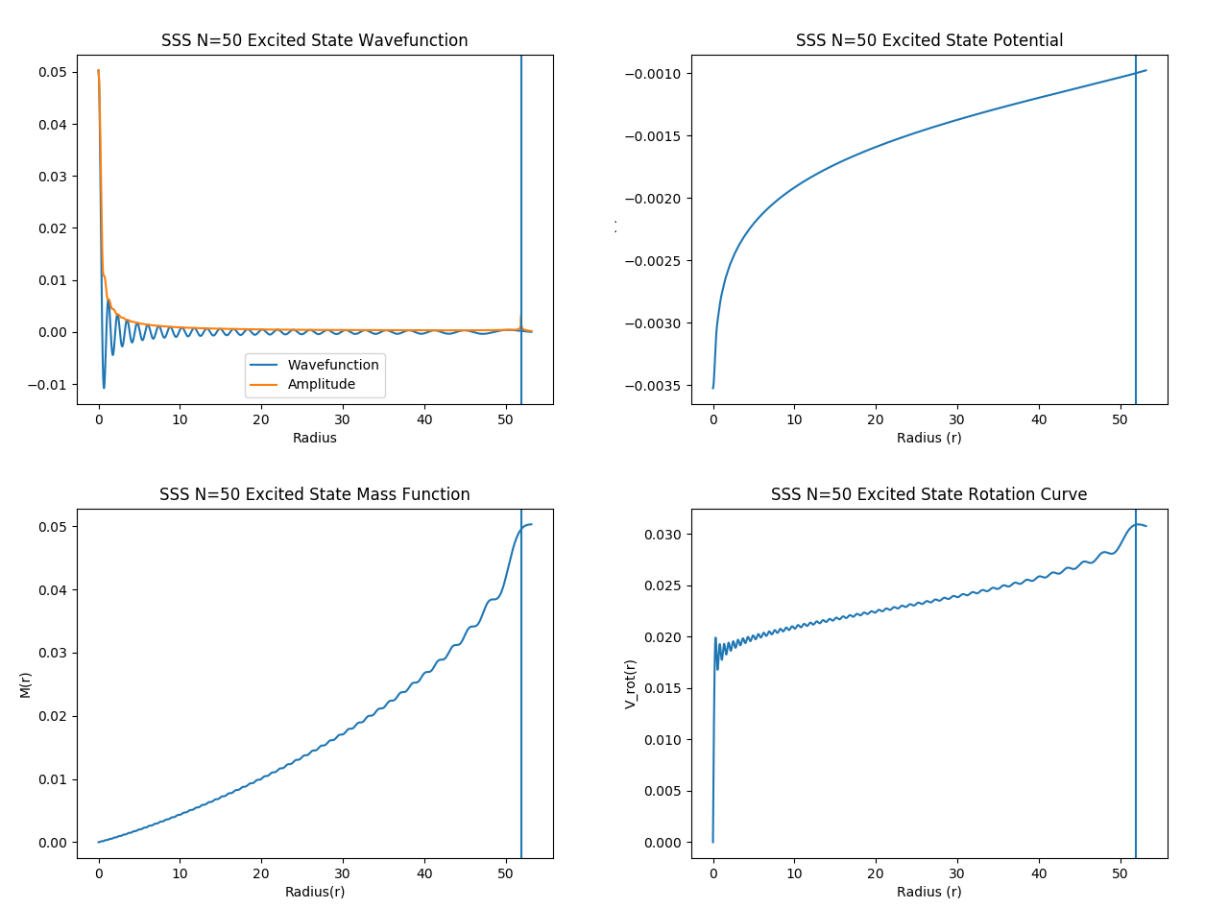}
    \caption{Plots rendered using units of $c=G=\hbar=1$, $m=100$, and a frequency of $\omega=99.9$.  Here we illustrate various features of the $N=50$ SSS state.  Top left:  Radial wave function along with the corresponding amplitude function.  The amplitude function matches the wave function in absolute value at each wave function extremum.  Top right:  Gravitational potential function, choosing the $V_\infty=0$ convention.  Bottom left: Mass function, depicted the total mass enclosed by a shell of radius $r$.  Bottom right: Circular velocity curve computed as $v_{circ} = \sqrt{\frac{M}{r}}$.  The value of the decay radius, $R_d$, is denoted by a vertical line in each plot.  }
    \label{fig:SSSComposite}
\end{figure}

We now list some relevant quantities: 

\begin{itemize}
    \item $m$: The $\psi$DM particle mass.
    \item $\phi(r,t)$: The total halo wave function.
    \item $\Phi(r)$: The radial part of the wave function.
    \item $V(r)$: The gravitational potential of the wave.
    \item $M(r)$: The mass enclosed by a sphere at radius $r$.
    \item $n$: The state excitation number
    \item $v_{circ}(r) = \sqrt{\frac{M(r)}{r}}$: The velocity of a perfectly circular orbit at radius $r$.  
    \item $\omega$: The static state frequency.  
    \item $\rho(r) = \Phi(r)^2$ : The dark matter density at radius $r$.
    \item $k(r)^2 = m-\omega+mV(r)$: The wave function spatial frequency.
    \item $A(r)^2 = \Phi(r)^2 + \frac{\Phi_r(r)^2}{k(r)^2}$: Amplitude function (see figure \ref{fig:SSSComposite}).  
    \item $R_d$: The decay radius, specified by $k(R_d)=0$.
    \item $\lambda(r) = \frac{2\pi}{k(r)}$: Wave function wavelength at radius $r$. \item $M_{tot} = \lim_{r\to\infty}M(r)$: The total dark matter mass.
    \item $V_{\infty} = \lim_{r\to\infty}V(r)$
    
\end{itemize}

For $(M(r),V(r),\Phi(r),\omega)$ to describe a physically reasonable dark matter halo, the values of $M_{tot}$ and $V_{\infty}$ must be finite.  Moreover, for solutions to be non-singular at $r=0$, both $M(0)=0$ and $\Phi_r(0)=0$ must be enforced.  The only such solutions are the SSS states as previously described in section \ref{Sec:SSSExcitedStates}.  The value of $V_{\infty}$ is a convention; we choose the standard convention of $V_{\infty}=0$.  However, it should be noted that the ability to shift the potential by a constant is a useful computational feature of this set of ODEs.  

The radial wavefunction has the symmetry of $\Psi(r)\to-\Psi(r)$.  Physically, this is understood by noting that the density, $\rho(r)$, depends on the square of the wave function.  We choose to use the convention that $\Psi(0)>0$ so that the wavefunction is always positive at the origin. 

Finite mass halos require that $\omega < m$ hold true.  This can be understood by considering the value of $k^2(r)$.  In the case that $\omega > m$, the value of $\lim_{r\to \infty}k^2(r)$ is always negative, and hence $\Phi(r)$ oscillates indefinitely, resulting in an infinite mass.  For $\Phi(r)$ to appropriately vanish at infinite distance, the value of $k^2(r)$ must eventually become positive so that $\Phi(r)$ displays exponential behaviors.  This can only be achieved if $\omega<m$, resulting in a $\Phi(r)$ which is initially oscillatory, reaches the decay radius, $R_d$, and then converts to an exponential decay for all further radii.  

The value of $\omega$ itself relates to total halo mass in the case that $V_{\infty} =0$.  Qualitatively, smaller values of $\omega$ correspond to larger total mass halos. One can thus adjust the total mass of the halo via changing $\omega$ and re-solving the system, or equivalently by applying the scaling relations described later in section \ref{Sec:ScalingRelations}.  Interestingly, as an aside, fixing the value of $\omega$ and taking $V_{\infty}=0$ results in an approximately linear relationship in the masses of the excited states, such that
\begin{equation}
\label{eq:Constantomega}
    M_{n,tot}(\omega) \approx (n+1)M_{0,tot}(\omega)  
\end{equation}

The amplitude function, $A(r)$, is devised to account for the fact that the wave function has oscillations of the scale of $\lambda(r)$.  We will later wish to track the average density scale of fluctuations in the DM halo.  To do so, we construct $A(r)$ with the formula in the above list.  This construction considers $\Phi(r)$ to be similar to functions with the following local behavior

\begin{equation}
\label{eq:OscillatoryFunction}
    f(R+r) = A(R)\sin(k(R)r+\delta(R))
\end{equation}
Then, one can extract an amplitude as
\begin{equation}
\label{eq:AmplitudeFunction}
    A(R)^2 = f(R+r)^2 + \frac{f_r(R+r)^2}{k(R)^2}
\end{equation}

The amplitude for $\Phi(r)$ then takes the form stated in the beginning of this section.  It should be noted that this definition of $A(r)$ is divergent at the decay radius due to the fact that $k(R_d)=0$, though for values of $r<R_d$, $A(r)$ approximates the extrema of $\Phi(r)$ well and is thus a useful tool.  For a graphical depiction of the amplitude, see figure (\ref{fig:SSSComposite}).

\subsection{Generic SF$\psi$DM Halos}
\label{Sec:GenericHalos}
A SF$\psi$DM halo is a complicated and turbulent structure.  High resolution simulations show that halos formed from multiple parent solitons generically display an inner soliton region and a turbulent outer region which eventually converges to an NFW-like profile \cite{SchiveCosmology,MoczGalaxy}.  Ultimately, we wish to describe these halos with generic solutions to the EKGEs, $\psi(\vec{r},t)$.  Omitting the non-linearity introduced by the Einstein equation, this would be a relatively simple task.  That is, if the spacetime metric is taken to be fixed and de-coupled from $\psi$, the Klein-Gordon-Equation which describes $\psi$ becomes linear.  This enables one to describe the halo as a linear combination of wavefunctions involving the spherical harmonics, with each wavefunction obeying the Klein-Gordon-Equation.  For the case of a complex scalar, this appears as follows:
\begin{equation}
\label{GeneralExpansion}
    \psi_{\omega,l}^j(\vec{r},t)=r^l Y_l^j(\theta,\phi)\Psi_{\omega,l}(r) e^{i\omega t}
\end{equation}
We note that this rather simple model was suggested for the case of a real scalar in \cite{BraySpirals} purposed towards generating spiral patterns in galaxies.  Given complete data of a halo, one could hypothetically find the best fit combination of these functions to describe the halo.  Working the other way around, exploring these solutions could give insight as to how galactic halos are shaped.  

As a first step to understanding the solutions in the above equation, we take the $j=0$ case of a single frequency $\omega$, resulting in the following form.    
\begin{equation}
\label{eq:SSSHarmonic}
    \psi(\vec{r},t) = \Psi(r)e^{i\omega t}
\end{equation}
This harmonic form produces the SSS wave functions for a complex scalar.  Even reintroducing the coupling to the spacetime metric, solutions to the fully SSS and coupled EKGEs have been categorized as described in the prior sections.

\subsection{SSS SF$\psi$DM Galaxies}
\label{Sec:SSSGalaxies}
So far we have discussed $\psi$DM halos in a purely DM-only context.  To extend this discussion we consider including additional sources of gravity via the potential function $V(r)$.  For simplicity, we will only consider the inclusion of static, spherically symmetric, external densities $\rho_{ext}(r)$.  To determine the potential $V_{ext}(r)$ to be included in the background of the DM halo, we assume the external component to approximately solve the Poisson Equation, thus
\begin{equation}
\label{eq:ExternalPoisson}
    \nabla^2V_{ext}\approx4\pi\rho_{ext}
\end{equation}

We then consider a slightly adjusted set of ODEs, which are most easily understood in the non-relativistic form corresponding to the PS system.
\begin{equation}
\label{eq:PSExtPsi}
    i \frac{\partial \psi}{\partial t} = -\frac{1}{2m}\nabla^2\psi+m(V+V_{ext})\psi
\end{equation}
\begin{equation}
\label{eq:PSExtV}
    \nabla^2(V+V_{ext}) = 4\pi (|\psi|^2+\rho_{ext})
\end{equation}

In terms of the SSS EKGEs, this is achieved in an approximate sense by making the following substitutions:
\begin{equation}
\label{eq:Vtot}
    V(r) \to V_{tot}(r) = V(r) + V_{ext}(r)
\end{equation}
\begin{equation}
\label{eq:Mtot}
    M(r) \to M_{tot}(r) = M(r) + M_{ext}(r)
\end{equation}
such that $M_{ext}(r) = \int_0^r (\rho(x)4\pi x^2 dx)$.
Since the external component obeys the Poisson equation, equations (\ref{eq:SSSPotential}) and (\ref{eq:SSSMass}) will be approximately unchanged.  However, the equation (\ref{eq:SSSSchrodinger}), now equation (\ref{eq:PSExtPsi}), for $\Psi(r)$ must now account for the externally applied potential.  

\begin{figure}
    \centering
    \includegraphics[height=3in,width=3.35in]{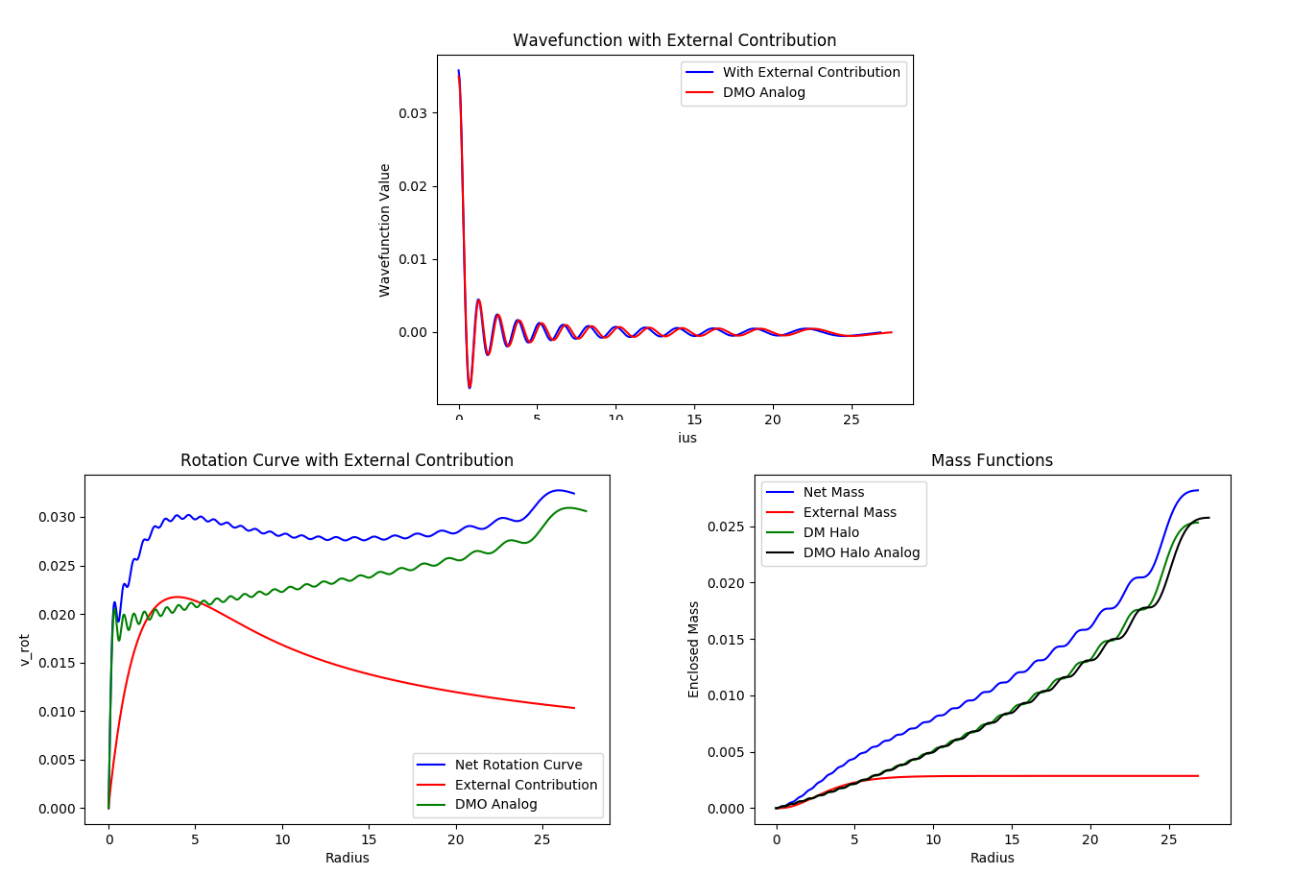}
    \caption{Figures rendered using units of $c=G=\hbar=1$, $m=100$, and frequency of $\omega=99.9$.  The effects of including additional matter contributions are displayed.  Here an external density of $\rho(r) = Ke^{-Cr}$ is used.  Total dark matter fractions are set to 90\%, with a fraction at the baryon half mass radius (see section \ref{Sec:ModelDetails}) of 50\%.  Top : The wave function is minimally changed, contracting slightly as a result of additional matter content in the background.  Bottom left: Including external components can greatly affect the overall shape of the rotation curve.  Sizeable contributions near the central region tend to flatten the overall rotation curve.  An analogous DM-only (DMO) halo is included for comparison.  Bottom right: Mass functions corresponding to the top two plots, again including a DMO analog.  A contraction of the overall galaxy as a result of including the external matter is evident, and can be seen by comparing the two DM halos.  }
    \label{fig:GalaxyComposite}
\end{figure}
Solutions to this new set of equations characterize SSS DM halos under the influence of SSS background density components.  These solutions can be thought of as analogous to the ones in the DMO setting, but with alterations in shape depending on the relative size and distribution of the included background density in comparison to that of the DM halo.  In most contexts, due to the relatively small ratio of baryonic mass to DM mass this will result in only small changes to the overall solution in comparison to the DM-only setting.  The overall rotation curve of the galaxy however, can drastically change due to the inclusion of the external density.  This feature is illustrated in figure (\ref{fig:GalaxyComposite}).  

\section{The Baryonic Tully-Fisher Relation, $\psi$DM Boundary Conditions and Scaling Relations}
\label{Sec:BTFRBoundaryCondsScalings}
\subsection{The Tully-Fisher Relation and $\psi$DM}
\label{Sec:TFRandWDM}
\begin{figure}
    \centering
    \includegraphics[height=2in,width=3in]{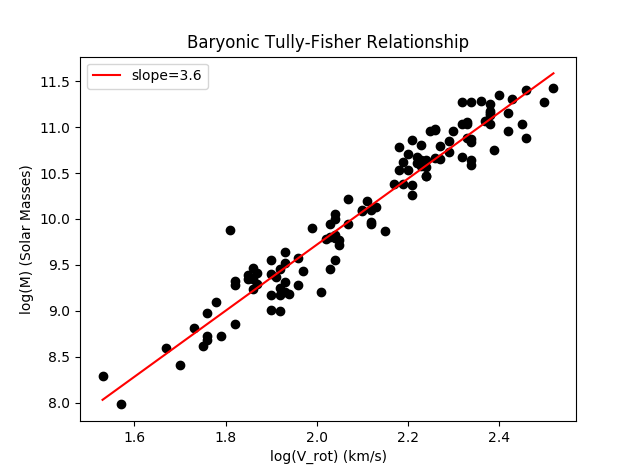}
    \caption{The Baryonic Tully Fisher Relation:  Here the BTFR is depicted as a power law relation between total baryonic mass and circular velocity.  Data retrieved from SPARC survey \cite{SPARCData} (ommitting error bars).  A light to mass ratio of $\Upsilon_*=0.5$ is assumed for the $[3.6]$ band.  Qualitatively, higher values of $\Upsilon_*$ result in steeper slopes on the Log-Log plot. }
    \label{fig:BTFR}
\end{figure}

The Tully-Fisher Relation (TFR), depicted in figure (\ref{fig:BTFR}) is an empirical relation regarding disk-like galaxies reported in 1977 \cite{TullyFisherOriginal}.  The original result stated that the total luminosity, $L$, of a galaxy was related to the width, $\delta$, of its observed 21cm spectral line in the following way, for some exponent, $x$:
\begin{equation}
\label{eq:LineWidth}
    L\propto\delta^x
\end{equation}
$\delta$ can then be related to the maximal rotational rate of the galaxy, $v_{m}$, implying a similar relation between $L$ and $v_{m}$.
\begin{equation}
\label{eq:LuminosityVelocity}
    L\propto v_m^x
\end{equation}
In practice, the TFR is used as a distance measure for spiral-type galaxies.  By measuring the line-width, or rotational velocity, the TFR can be used to derive the luminosity of the galaxy.  This luminosity can then be compared to the apparent brightness of the galaxy, which in turn directly relates to its distance measure.  For this reason, the TFR is often included as a rung in the cosmic distance ladder.  

In 2000, it was observed by McGaugh that the TFR failed to describe many low surface brightness (LSB) galaxies \cite{LowSurfaceBrightness}.  This was explained by the fact that LSB galaxies have a higher fraction of their mass contained in gas, as opposed to in luminous stars.  As a result of their high gas fractions, LSB galaxies have rotational velocities greater that what is predicted by the TFR if only the luminous stellar mass is measured.  It was proposed that this issue could be resolved by measuring the \textit{total baryonic mass}, $M_b$, and that instead
\begin{equation}
\label{eq:MassVelocity}
    M_b\propto v_m^x
\end{equation}
This relation is referred to as the \textit{Baryonic} Tully-Fisher Relation (BTFR).  This relation reveals a linear relation between $\log(M_b)$ and $\log(v_m)$ with a slope of $x$. The actual value of $x$ is a topic worthy of discussion.  In fact, the slope strongly depends on the assumed stellar-light-to-mass-ratio, $\Upsilon_*$, for the galaxy sample used. The correct value for this ratio is still debated, and changes depending on galaxy sample; in fact, some studies suggest that $\Upsilon_*$ may vary within individual galaxy samples \cite{BellStellarLTMR}.  We compare to the SPARC survey which chooses the 2.2$\mu$m band for rotation curve data under the reasoning that $\Upsilon_*$ is predicted to be nearly constant in this band for a range of galaxy types.  In this band, typical values of $x$ lie in the range of 3 to 4 \cite{McGaughTFRScatter}. More details of the SPARC survey are provided in section \ref{Sec:DataDetails}.

Though the BTFR relation describes the baryonic mass component of galaxies, the circular velocity of the galaxy is a function of the \textit{total} mass of that galaxy, $M(r)$.  In the non-relativistic case, circular velocities are computed as: 
\begin{equation}
\label{eq:CircularVelocity}
    v_m = \sqrt{\frac{M(R_m)}{R_m}}
\end{equation}
where here $R_m$ is the radius of the maximal rotational velocity as measured from the axis of rotation.  This allows one to use the BTFR to make inferences regarding the dark matter content of galaxies.  

\begin{figure}
    \centering
    \includegraphics[height=1.75in,width=3.35in]{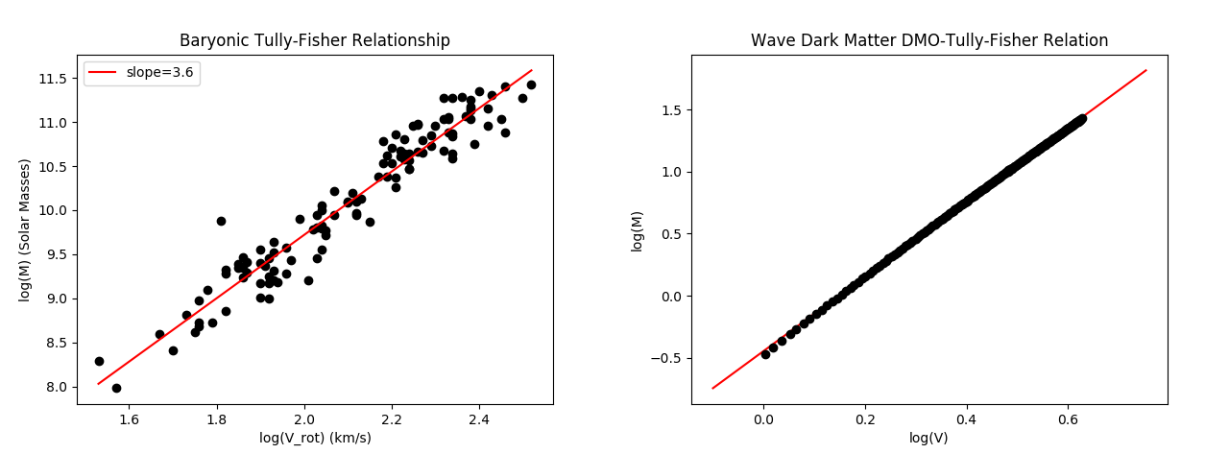}
    \caption{$\Psi$DM Tully-Fisher Relation.  Left: BTFR as reported from SPARC.  Right:DMO relation for 200 $\psi$DM excited states, scaled in computational units to have $\Phi(R_d)=1$ to reproduce boundary conditions as in \cite{GoetzThesis}.  Slope of the DMO relation asymptotically approaches $3.4$ for large values of excitation number, but deviates for small $N$.}
    \label{fig:GoetzTFR}
\end{figure}

It was noted in \cite{GoetzThesis} that a relation similar to the BTFR could be found in DM-only simulations of SSS excited states to the EKGEs.  It was shown that the SSS excited states could reproduce a relation between the dark matter mass of those states, $M_{dm}$, and their maximum circular velocity.
\begin{equation}
\label{eq:DMTFR}
    M_{dm}\propto v_m^y
\end{equation}
Moreover, upon imposing the appropriate boundary conditions to the excited states, the value of $y$ was shown to lie in the generally accepted range between 3 and 4, with analytical arguments indicating a value of $y\approx3.4$ \cite{GoetzThesis}.  An example of this relation is given in figure (\ref{fig:GoetzTFR}).  This observation suggests that the SSS excited states of $\psi$DM could provide a mechanism for generating the BTFR.  In the following sections we suggest and detail a method to determine what values of the $\psi$DM mass parameter are compatible with the observed BTFR.  

\begin{figure}
    \centering
    \includegraphics[height=3in,width=3.35in]{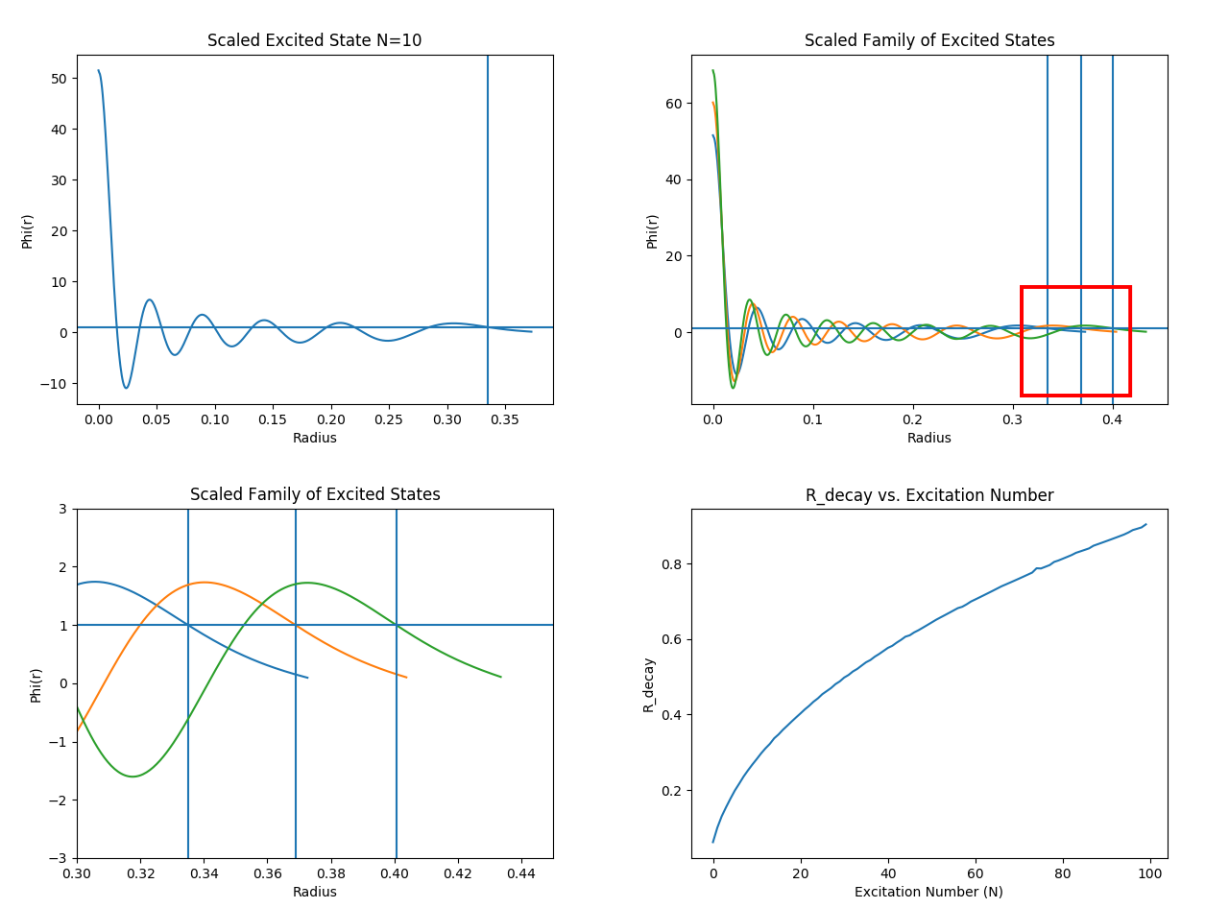}
    \caption{Top Left: An excited state of order $N=10$ scaled to have $\Phi(R_d)=1$.  Vertical line denotes the location of the decay radius, $R_d$.  Horizontal line denotes the boundary value of $1$.  Top Right: Three successive excited states subject to the same boundary conditions. Again, vertical lines denote the decay radius.  Bottom left: A zoomed in image of the boxed region from the top right image, showing the region at which the boundary conditions are applied for each state.  Bottom right: The decay radius as a function of $N$.  Empirically, $R_d\sim \sqrt{N}$.  Qualitatively, as $N$ increases, the values of $R_d$ become more closely spaced.  }
    \label{fig:GoetzBC}
\end{figure}

The family of halos produced in \cite{GoetzThesis} can be understood as the result of imposing a particular boundary condition on the the SSS EKGEs.  Explicitly, given an $n$th excited radial wave function, $\Phi_n(r)$, with  decay radius of $R_{d}$, this boundary condition can be stated as

\begin{equation}
\label{eq:DecayBoundary}
\Psi_n(R_d,n) = \bar{\Psi}
\end{equation}
where $\bar{\Psi}$ is the same constant for all values of $n$.  An example of such a family is given in figure (\ref{fig:GoetzBC}).  An important feature of this boundary condition is that it is always applied at the decay radius $R_d$, which is defined by the property that the wave number, $k_n(R_d,n) = 0$.  In other words, choosing the boundary point to be $R_d$ prescribes the same value of $k(R_d,n)$ at the boundary \textit{before the value of $\Psi$ is ever specified}.  Qualitatively, this boundary condition could be interpreted as fixing a density scale in the far regions of the wave function, choosing the decay radius as the scaling point. 

As an aside, we emphasize to the reader that the Tully-Fisher-like trend displayed in figure \ref{fig:GoetzTFR} is an unexpected and rather remarkable feature of $\psi$DM excited states.  This feature suggests a theoretical mechanism for producing the BTFR which has not been displayed in other models of DM; the fact that SSS excited states are able to reproduce such an appropriate relation is therefore extremely interesting. It will be our goal in the following sections to adapt this feature to produce simulations compatible with the observed BTFR.   

\subsection{On Halo Boundary Conditions}
\label{Sec:HaloBoundaryConds}
Choosing solutions to the EKGEs which reproduce families of galaxies similar to those observed is a computationally intensive process.  It is observed in \cite{GoetzThesis} that physically reasonable halos can have vastly different global properties depending on what type of boundary condition the halos satisfy.  Furthermore, choosing a boundary condition at the decay radius as in (\ref{eq:DecayBoundary}) was shown to reproduce SSS solutions which are consistent with a Tully-Fisher-Like relationship.  We wish to choose a set of physically motivated boundary conditions which reproduce this property.  

\begin{figure}
    \centering
    \includegraphics[height=3in,width=3.35in]{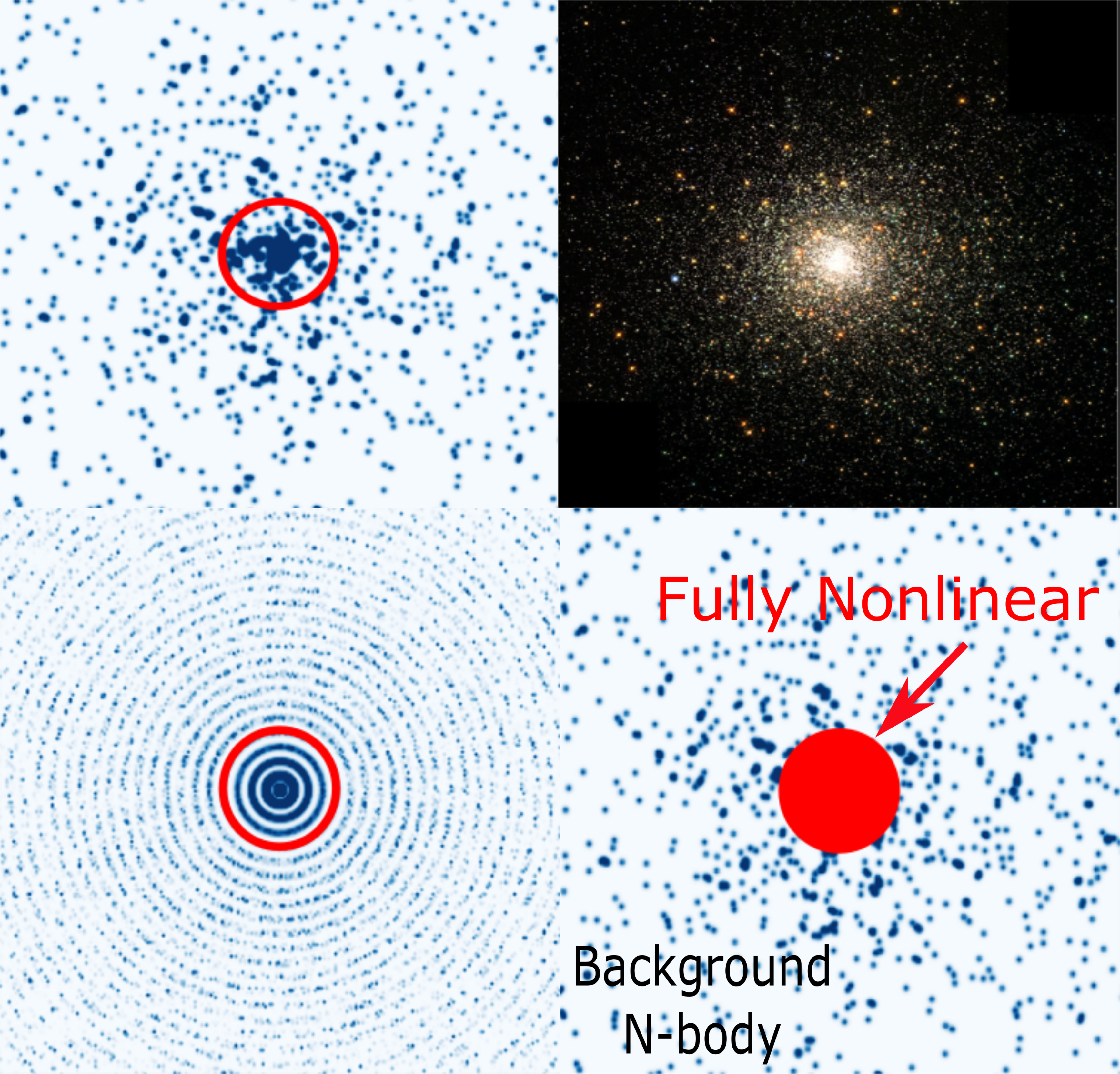}
    \caption{Top left: Many identical solitons placed uniformly in radius and randomly in angle to replicate an N-body problem. Densities are summed in regions with sufficient overlap of states, and a saturation level is applied.  The red circle denotes a boundary between an inner region with high overlap and an outer region with low overlap.  Top right:  A globular cluster of stars \cite{GlobularCluster} as a comparative visual of an N-body problem.  Bottom left: The same as the top left image, but placed uniformly in angle as well as radius.  Bottom right:  An interpretation of $\psi$DM halos.  A central region described by a nonlinear wave, and an outer region with many constituent solitons interacting in an N-body fashion. }
    \label{fig:BoundaryConditionNBody}
\end{figure}

In figure \ref{fig:BoundaryConditionNBody}, we depict a galactic halo formed by many $\psi$DM solitons.  This image will serve merely as the \textit{motivation} for our proposed physical boundary conditions.  In the top left panel of the image we depict an N-body system comprised of many solitons; this is compared to an N-body system of stars forming a globular cluster.  Though the two images should occur at vastly different physical scales, we only wish to compare the qualitative features.  In the central regions of the system, there is a large amount of overlap (and therefore interference) between the constituent solitons.  In the outer regions, the solitons are much more separated, and appear as point particles.  We suggest there to be a boundary between these regions, depicted by the red circle, at which the halo transitions between these two regions.  

The bottom left panel of figure \ref{fig:BoundaryConditionNBody} depicts the same system, but in a spherically symmetric setting.  This corresponds to the approximation of the halo by a spherically symmetric function as in equation (\ref{eq:SSSHarmonic}).  At the indicated boundary, the wavelength scale of the individual solitons is large enough in comparison to their number density that they begin to overlap.  We suggest that this boundary is dependent upon such a wavelength scale, $\lambda_{DM}$.  Moreover, this wavelength scale must be matched with the density scale $\rho_{DM}$, of the constituent solitons.  This gives us our first conception of a galactic boundary condition.  For some generic boundary value, $R$, the halo wave function satisfies
\begin{equation}
\label{eq:BoundaryConds1}
(\lambda(R),\rho(R)) = (\lambda_{DM},\rho_{DM})
\end{equation}
Then, to account for the fact that that density displays fluctuations of the size $\rho_{DM}$, we choose to consider the density amplitude $A_{DM}$ as opposed to $\rho_{DM}$ itself.  This then sets a scale for the maximal size of the fluctuations themselves.
\begin{equation}
\label{eq:BoundaryConds2}
(\lambda(R),A(R)) = (\lambda_{DM},A_{DM})
\end{equation}

In the final, bottom right panel of figure (\ref{fig:BoundaryConditionNBody}), we suggest a more complete model for $\psi$DM halos.  This can be understood in the following way:  In the outer regions of the halo, its constituent solitons are separated greatly enough that they behave as point particles, and thus form an N-body system.  Eventually, at regions near the described boundary, the solitons begin to overlap and interfere significantly, and can no longer be individually resolved.  The center region then, is described by a generic solution to the non-linear EKGEs.  

We finally note that the interpretation from figure (\ref{fig:BoundaryConditionNBody}) is rather simplified.  In fact, simulations of bottom-up formed halos have not reported an outer region populated with solitons, but instead a turbulent region filled with field fluctuations.  The fluctuations themselves occur on the deBroglie wavelength scale of the DM halo, and are comparable in width to the central core \cite{MoczGalaxy,CoreHaloRelation}.  Such fluctuations are sometimes deemed ``quasiparticles," and are not persistent objects like solitons.  However, these quasiparticles do survive for a short time period, and in this time are subject to the same gravitational accelerations that stable particles would be subject to.  We therefore consider the possibility that for short instants in time, the turbulent galactic halo appears similarly to the interpretation in figure (\ref{fig:BoundaryConditionNBody}), populated by $\psi$DM particles or quasiparticles.  

As a final note regarding this boundary condition, we emphasize that the soliton ground state is the only stable attractor solution to the PSEs.  It is thus reasonable to suppose that $\psi$DM structures tend to evolve towards configurations favorable to forming such solitons.  If the fluctuations of the outer halo are able to reach the kinetic regime as in \cite{Condensation}, then one may expect such soliton condensation to occur.  This begs the question ``Does soliton condensation occur in galaxies, and if so does it have a preferred mass scale?"  Moreover, one could suppose that such a mass scale is a result of the halo formation process, and is related to its parent solitons.  This interpretation could have implications for the formation of solitons in the early universe in that it may suggest an average size at which the condensation of $\psi$DM solitons occurs.  These solitons could then merge in a bottom-up fashion, carrying over the properties of their mass and length scale to the resultant halo.     

\subsection{Scaling Relationships and Invariant Quantities}
\label{Sec:ScalingRelations}
We now detail a set of relations which will be useful in solving such boundary problems. A useful feature of solutions to the PSEs is that they admit a set of exact scaling relations (and hence the EKGEs admit such relations in an approximate sense).  Firstly, denote an $n$th state solution as $(M_n(r),V_n(r),F_n(r),\omega_n)$.  Now, given any two real numbers, $\alpha$ and $\beta$, another $n$th state solution, $(\bar{M_n}(\bar{r}),\bar{V_n}(\bar{r}),\bar{F_n}(\bar{r}),\bar{\omega_n})$, can be generated with the following relation:  

\begin{equation}
\label{eq:rscaling}
\bar{r} = \alpha^{-1}\beta^{-1} r
\end{equation}
\begin{equation}
\label{eq:Mscaling}
\bar{M}(\bar{r}) = \alpha\beta^{-3}M(r)
\end{equation}
\begin{equation}
\label{eq:Vscaling}
\bar{V}(\bar{r}) = \alpha^2\beta^{-2}V(r)
\end{equation}
\begin{equation}
\label{eq:Psiscaling}
\bar{\Psi}(\bar{r}) = \alpha^2\Psi(r)
\end{equation}
\begin{equation}
\label{eq:mscaling}
\bar{m} = \beta^2m
\end{equation}
\begin{equation}
\label{eq:omegascaling}
\bar{m}-\bar{\omega} = \alpha^2(m-\omega)
\end{equation}

These relations can be further used to compute quantities which are \textit{scale invariant} for the PS solutions; this property will later be useful in generating appropriately scaled families of excited state solutions.  For instance, the Mass-Radius quantity, 
\begin{equation}
\label{eq:MassRadius}
    \mu(r) = M(r)r=\bar{M}(\bar{r})\bar{r}
\end{equation}
is scale invariant assuming a constant value of $m$ (hence for $\beta =1$). If one parameterizes $\mu(r)$ in terms of $r'=\frac{r}{R_d}$, then the result is a function $\mu(r')$ which is unchanged upon applying scalings.  Moreover, any product of characteristic mass and radius values will yield an invariant quantity.  For instance, taking the total mass of the state as $M_{tot}$ and the radius containing half of that total mass, $R_{1/2}$ we can form an invariant which is unique to each $n$th order state, namely 
\begin{equation}
\label{eq:MassHalfRadius}
    I_n = M_{tot,n}R_{1/2,n}
\end{equation}
In the case of the ground state, this reproduces the relation found in equation \ref{eq:GSMass}.  Qualitatively then, for a given $n$th state, a higher mass corresponds to a smaller radius; more massive states are more contracted.  Moreover, once the mass of the halo is specified, its length scale is predetermined and vice versa.

The scaling relationships above apply to the DM-only PSEs as presented in equations (\ref{eq:SSSSchrodinger}),(\ref{eq:SSSPotential}) and (\ref{eq:SSSMass}).  Later, we will need to consider cases with external potentials and choose scales appropriate to match the BTFR. In doing so, one must apply the correct scaling to the external component in order to ensure that the re-scaled result is indeed a solution of the PSEs.  This can be achieved as long as the same types of relations are applied to the external component.  That is, we must re-scale the external potential, and mass functions in exactly the same manner as listed for the DM functions.  Therefore, once the relative ratios of the DM and external components are chosen, the system of ODEs may first be solved at any convenient computational scale and then the scaling relations can be applied to re-scale to a desired sizing (a particular total mass for instance).

\subsection{Amplitude-Wavelength Boundary Conditions}
\label{Sec:AmpWaveengthBoundary}
In the following sections we solve a particular boundary problem in order to generate a family of $\psi$DM galaxies which fit the BTFR.  This boundary problem can be thought of as imposing the following requirements, denoting an $n$th state with a subscript $n$.
\begin{equation}
\label{eq:Adm}
    A_n(R_n) = A_{DM}
\end{equation}
\begin{equation}
\label{eq:lambdadm}
    \lambda_n(R_n) = \lambda_{DM}
\end{equation}

\begin{figure}
    \centering
    \includegraphics[height=1.75in,width=3.35in]{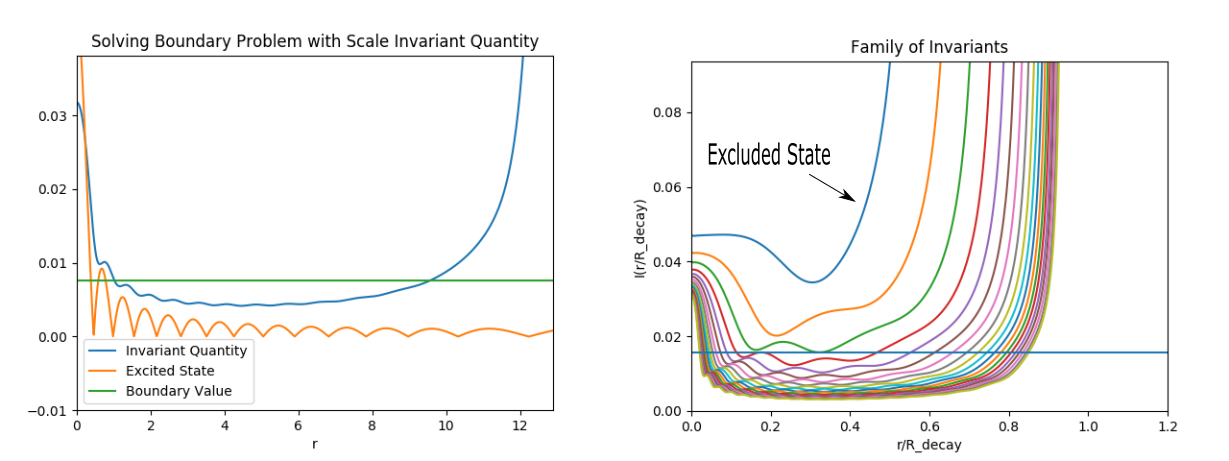}
    
    \caption{Solving Boundary Problems with scale invariants: Left: For each state, we construct a scale invariant function, $I(r)$ relevant to the boundary value.  Then, choosing a boundary value, shown in the figure by a constant value of the invariant, we find the intersection with $I(r)$.  The resulting intersection is a characteristic point at which the desired boundary problem can be solved.  Solutions are then rescaled based off of this point.  Right: Family of scaling invariant functions for each excited state.  Each successive value of $n$ displays the invariant bounded above by the previous value of $n$.  Invariants plotted in units of $\frac{r}{R_{d}}$ for each state.  States displaying no intersection cannot satisfy the boundary condition, and can therefore be excluded.}
    \label{fig:BoundaryInvariant}
\end{figure}

In other words, we wish to apply the same amplitude and wavelength scale to each $n$th state at some general values of radius, $R_n$.  Though this problem can be solved rather robustly using shooting problem methods, making use of the scaling relations can simplify the problem and greatly increase throughput.  Due to the scaling properties of the PSEs, only certain products of $\lambda_{DM}$ and $A_{DM}$ are attainable at a generic boundary.  This can be understood as follows:  as a result of the scaling relations, the following value forms a scale invariant function characteristic to each $n$th state.
\begin{equation}
\label{eq:amplitudelambda}
    I_n(X) = A_n(X)\lambda_n(X)^2
\end{equation}
Here we have used the definition of $X=\frac{r}{R_d}$.  At each characteristic value of radius, $X$, an $n$th state has a predetermined product $I_n(X)$.  Therefore, if one supposes a value for a boundary, $I_{DM} = A_{DM}\lambda_{DM}^2$, the intersection of $I_{DM}$ with $I(X)$ picks out the value of $X$ at which the excited state can achieve the prescribed product.  We will denote this intersection as $X_n$, highlighting that this intersection occurs at a different characteristic radius for each state.  We depict this intersection on the left in figure (\ref{fig:BoundaryInvariant}).

Once the point $X_n$ is found for a given state, the values of $\lambda(X_n)$ and $A(X_n)$ have achieved the appropriate product for the boundary condition at $X_n$, though the individual values of $\lambda(X_n)$ and $A(X_n)$ may not match their individually desired values.  However, a simple application of the PSE scalings will fix this issue.  That is, choose a scaling parameter $\alpha_n$ (taking $\beta=1$) such that $\alpha_n^{-1}\lambda(X_n)=\lambda_{DM}$.  This scaling will also enforce the appropriate value for $A(X_n)$ as a result of the scale invariance of $A(X_n)\lambda(X_n)^2$.  The resulting re-scaled state, then may be reparameterised by the physical radius $R=XR_d$ and will display the quality that $(A(R_n),\lambda(R_n))=(A_{DM},\lambda_{DM})$ for the boundary radii $R_n = X_nR_d$.

As depicted in figure (\ref{fig:BoundaryInvariant}), $I(r)$ displays a divergence at $X=1$.  This has important implications for the solutions of this boundary problem.  Seen in figure (\ref{fig:BoundaryInvariant}), simulations display that for each successive value of $n$, $I_n(X)$ is bounded above by the previous value of $n$.  This results in the feature that $\lim_{n\to\infty}X_n=1$.  In other words, as $n \to\infty$, the boundary problem is solved nearer and nearer to the decay radius.  In this sense, the actual boundary value of of $I_{DM}$ becomes nearly irrelevant towards the result of the boundary problem in the limit of large $n$, almost always resulting in a re-scaling of the wave function at the decay radius.  This is quite similar to the boundary condition investigated in \cite{GoetzThesis}, and in fact converges to the same result as $n\to \infty$.  It would then be expected that for sufficiently large excitation number, this boundary problem reproduces a TF-like relationship with a slope of $3.4$ as in \cite{GoetzThesis}.  We do however note that this method allows for fitting at radii $r<R_d$, which we consider to be an improvement.  Moreover, this method highlights the possibility of solving other types of boundary problems through a similar use of scale invariant function.

The fact that the limiting behavior of this boundary problem is independent of the value of $I_{DM}$ is rather convenient.  This implies that the resulting family of excited states only depends on the chosen scaling.  Once $I_{DM}$ is set, the scaling value, $\alpha_n$, for each excited state will be determined by the chosen value of $A_{DM}$ (equivalently one could choose $\lambda_{DM}$).  Each value of $A_{DM}$ corresponds to a unique family of halos.  Later, when computing a fit to the BTFR, this will result in the feature that the family of scalings which produces the best fit is unique and independent of $I_{DM}$.  That is, even when choosing a different $I_{DM}$, the $A_{DM}$ which provides the best fit will produce the same values of $\alpha_n$, and therefore the same family of halos.     

Another observation, indicated by the two intersection points in the left of figure (\ref{fig:BoundaryInvariant}), is that there are two possible families of solutions to the proposed boundary problem.  The first family, indicated by the leftmost intersection point, results in an unappealing property in regards to matching the BTFR.  In fact, this family will result in a negatively sloped relation as opposed to a positively sloped one as desired.  This can be understood by again considering the fact that each successive $I_n(r)$ is bounded above by the previous value of $n$.  The resulting family will therefore have decreasing values of the galactic boundary $R_n$ for an increasing galactic mass, conflicting data from observations.  The second family of solutions, indicated by the rightmost intersection, displays the opposite quality, having increasing values of the boundary $R_n$, and resulting in positively sloped relations as observed.  

One last observation regarding this boundary problem method is that it allows one to exclude states based on the desired value of $I_{DM}$.  As seen in figure (\ref{fig:BoundaryInvariant}), each function $I_n(\frac{r}{R_{decay,n}})$ displays a global minimum.  If $I_{DM}$ is chosen to be lesser than this minimum, then the state corresponding to $I_n$ cannot achieve the boundary value at any real value of radius.  Moreover, this excludes all other states of lower order.  That is, if $N$ corresponds to the highest order state such that $\min(I_{N}(r))>I_{DM}$, then all other states with $n<N$ are excluded from the family of solutions, as they will never intersect the boundary value. 

\section{Simulation of the BTFR}
\label{Sec:SimulationBTFR}
\subsection{Data} 
\label{Sec:DataDetails}
We choose to model the BTFR based on the data reported by the Spitzer Photometry \& Accurate Rotation Curves (SPARC) survey \cite{SPARCData}.  SPARC is a set of 175 extended H1 rotation curves collected from a set of several large surveys, including H1 observations from the Westerbork Synthesis Ratio Telescope (WRST), the Very Large Array (VLA), the Australia Telescope Compact Array (ATCA), and the Giant Metrewave Radio Telescope (GMRT).  The 175 galaxies are composed of both spirals and irregulars of 5 dex of stellar masses, 3 dex of surface brightnesses, and a variety of gas fractions. These observations were paired with infrared 3.6$\mu$m images from the Spitzer archive, detailing the stellar mass distributions for each sample.  Further details of the total SPARC sample are specified and presented in \cite{SPARCData}.

Specifically, for the BTFR data, we use the galaxy sample prepared in \cite{McGaughTFRScatter}, which contains 118 samples chosen based on having flat rotation curves and small angles of inclination. The total baryonic mass of each galaxy is reported as
$$M_b= M_g + \Upsilon_*L_{[3.6]}$$
for a gas mass $M_g$, a stellar mass-to-light ratio of $\Upsilon_*$ and 3.6$\mu$m luminosity $L_{[3.6]}$.  $\Upsilon_*$  can vary from galaxy to galaxy, depending on stellar content.  However, given that the ratio is expected to be nearly constant in the 3.6$\mu$m band, we do not consider the effects of its variations in this analysis.  It should be noted that varying the value of $\Upsilon_*$ will change the slope, the intercept, and the scatter of the BTFR, as demonstrated by \cite{McGaughTFRScatter}.  We choose to assume a constant value of $\Upsilon_* = 0.5$, which minimizes the intrinsic scatter of the data sample. 

\begin{figure}
    \centering
    \includegraphics[height=3in,width = 3.35in]{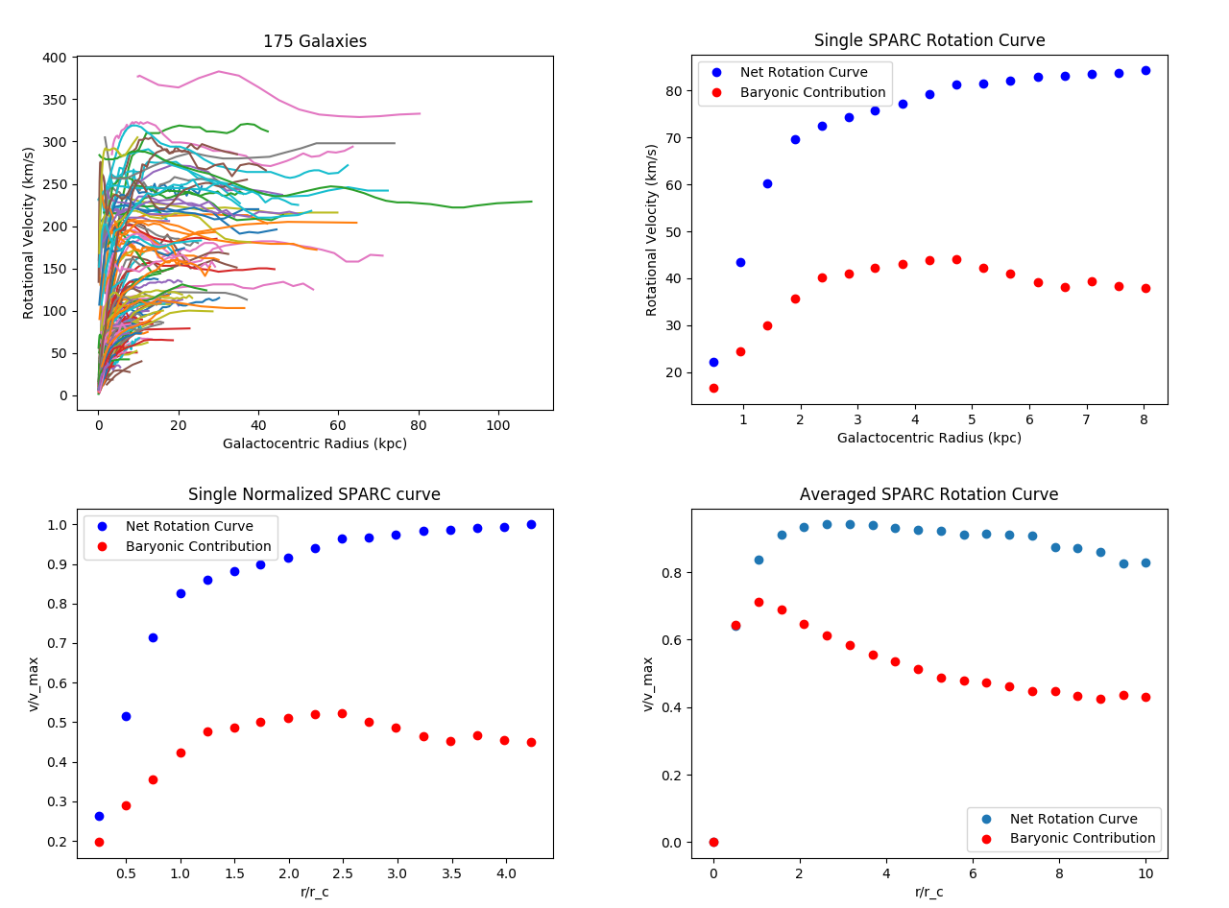}
    \caption{Top Left: All 175 rotation curves from the SPARC sample.  Top Right: Data points from a single rotation curve of the SPARC sample.  Bottom Left: The same single rotation curve, but with the radii scaled to $\frac{r}{r_c}$ ($r_c$ defined as in equation \ref{eq:CoreHalo}) and with velocities scaled to have their maximum value as $v_{max}=1$. Bottom Right: A normalized average of all SPARC rotation curves.  Averages are computed by normalizing each curve as in the bottom left figure, generating a spline for each normalized curve, and taking an equal weight average of all resulting splines for each value of $r$.  This is similar to the averaging procedure of \cite{AverageRotationCurve}.  If an individual sample lacks data at a given radius, it is not included in the average for that radius.  The baryonic contributions are included, and correspond to $v_{b}(r) = \sqrt(\frac{M_b(r)}{r})$}.
    \label{fig:SPARCData}
\end{figure}

An illustration of the SPARC rotation curves is given in \ref{fig:SPARCData}.  The rotational velocity corresponding to the BTFR velocity for the SPARC data is taken at the flat part of the rotation curve as defined in \cite{McGaughTFRScatter}.  For consistency, we sample the rotation curve of each simulation at $R_{circ} = 0.5R_{decay}$, which effectively selects the flat part of nearly all rotation curves produced by our $\psi$DM model. The fully prepared BTFR sample is represented as a log-log relation in figure (\ref{fig:BTFR}).

\subsection{Modeling Details}
\label{Sec:ModelDetails}
We now consider detail the model used to produce a family of $\psi$DM galaxies which will be used to reproduce the BTFR. Firstly we construct the SSS DM-only excited states of order 1-200 at a constant value of $\omega=.999\Upsilon$ (note that this value of $\omega$ is chosen for convenience, and will later be re-scaled), and then include an external contribution through a density $\rho_{ext}(r)$ as described in section \ref{Sec:SSSGalaxies}.  Computing the SSS DM-only states is fairly involved in a computational sense; we describe a method to attain these states in appendix I.  For the external component, we neglect all dynamical baryonic processes, and only consider the gravitational effects of the component.  Specifically we choose to use an exponential profile as a simple description of this component.
\begin{equation}
\label{eq:rhoext}
    \rho_{ext}(r) = Ke^{-Cr}
\end{equation}
Assuming that this component obeys the Poisson equation, this density corresponds to a gravitational potential of

\begin{equation}
\label{eq:Vext}
V_{ext}(r) = \frac{-K}{C^3}\left(\frac{2}{r}\left(1-e^{-Cr}\right)-Ce^{-Cr}\right)
\end{equation}
This potential must be included when solving the EKGEs.  We achieve this by defining a continuation parameter $\tau$.  That is, in the solving routine, we replace the DM-only potential $V(r)$ with a total gravitational potential
\begin{equation}
\label{eq:VTot}
    V_{total}(r) =V(r) + \tau V_{ext}(r)
\end{equation}

and then tune $\tau$ slowly from 0 to 1 in order to transition from the DM-only solution to a solution which fully includes the background potential.  This method is an adaptation of a method described in \cite{MarzuolaExternal}, and is fairly robust for small steps in $\tau$. 

The values of $K$, and $C$ are chosen to determine the total mass and length scale of the external disk.  Specifically
\begin{equation}
\label{eq:Mexttot}
M_{ext,tot} = \frac{8\pi K}{C^3}
\end{equation}
\begin{equation}
\label{eq:Rhalf}
    R_{b} \approx \frac{2.67}{C}
\end{equation}
where $R_{b}$ is the radius at which half of the external component is contained.  
In order to choose values for these parameters, we consider the dark matter fractions of the resulting halo-disk pair.  We choose to fix the total dark matter fraction, $f_{tot}=\frac{M_{DM,tot}}{M_{tot}},$ and the dark matter fraction at $R_{b}$, $f_{b}$.  This can be achieved by solving a shooting problem for both $K$ and $C$.  Multiple pairings of $f_{tot}$ and $f_{b}$ are considered, using values of $f_{tot}\in(0.7,0.9)$ and $f_{b}\in(0.5,0.9)$, which are consistent with those reported from IllustrisTNG simulations\cite{DarkMatterFractions}. 

Once halo-disk solutions are generated for the excited states, we impose the boundary conditions described in section \ref{Sec:BTFRBoundaryCondsScalings} on each state by choosing values of $\lambda_{DM}$ and $A_{DM}$.  Specifically, we choose values compatible with $\lambda_{DM}^2A_{DM}=0.4$ for the boundary problem so that no excited states are excluded from the resulting family\footnote{This is using our geometric computational units of $m=100$}.  Data corresponding to the simulated BTFR is then computed by extracting the total mass of the disk component of each solution, as well as the circular velocity at the flat part of the rotation curve (described in figure (\ref{fig:RotationFamily})).  We then adjust the set value of $A_{DM}$ so as to find the best fitting with respect to the SPARC data.  Finally, we repeat this process using multiple values of the $\psi$DM particle mass, $m\in(10^{-24}eV,10^{-21}eV)$.

\subsection{Results}
\label{Sec:Results}
\begin{figure}
  \centering

\includegraphics[height=3in,width=3.35in]{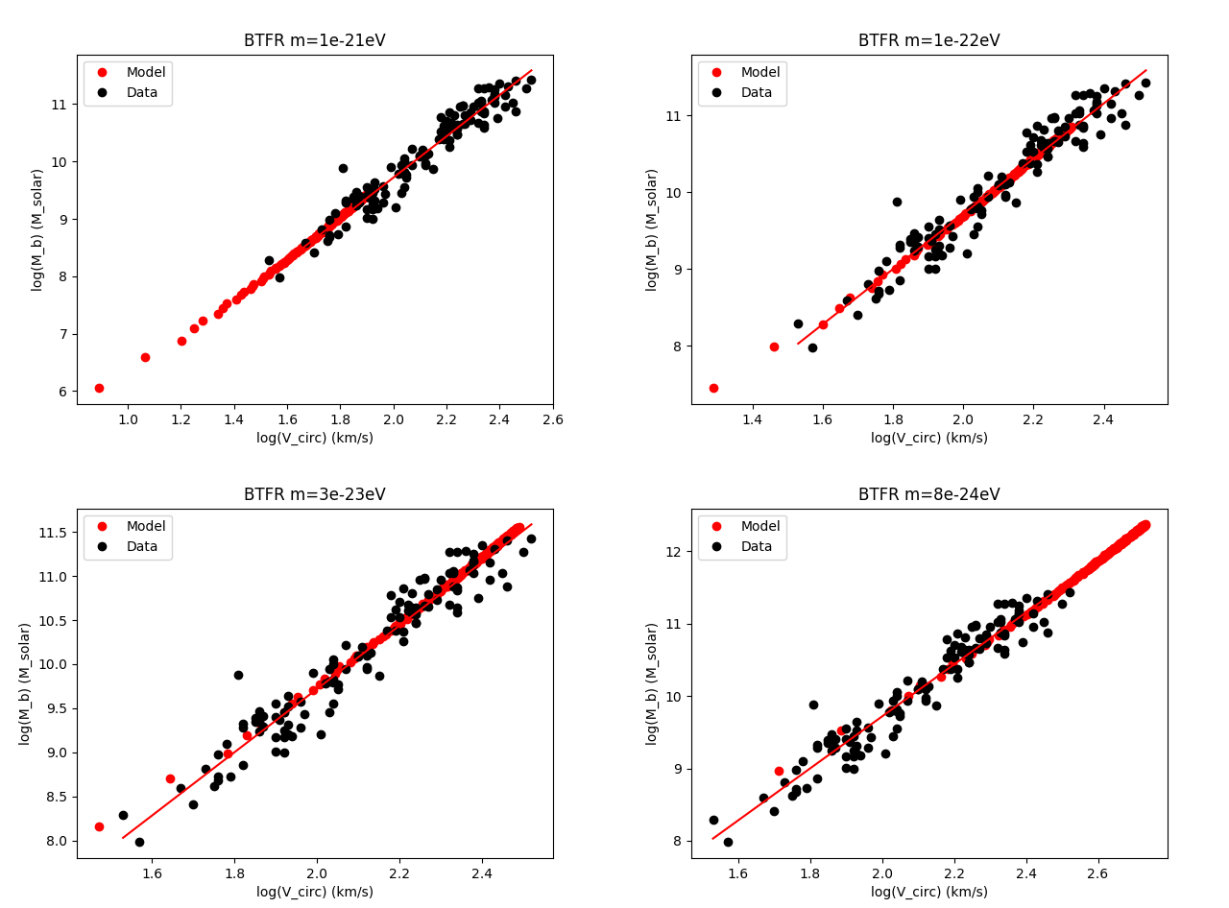}

\caption{Multiple simulated fits to the BTFR.  Total dark matter fractions of 0.9 are used for these plots.  SSS states of $n=0-200$ are included.  As seen, the value of the mass parameter affects the positioning of the SSS states in regards to the BTFR.  Higher values of $m$ correspond to smaller galaxies overall, shifting the simulated BTFR towards lower masses.  For sufficiently small masses, the simulated fit does not overlap the low mass end of the BTFR.  Choosing states $n\geq 0$ implies $m\geq10^{-23}eV$ while choosing $n\geq3$ implies $m\geq10^{-22}eV$. Including more states $n\geq200$ will extend the simulated fits to include higher galactic mass.}  
\label{fig:TFComposite}
\end{figure}

\begin{figure}
\centering

\includegraphics[height=1.5in,width=3.35in]{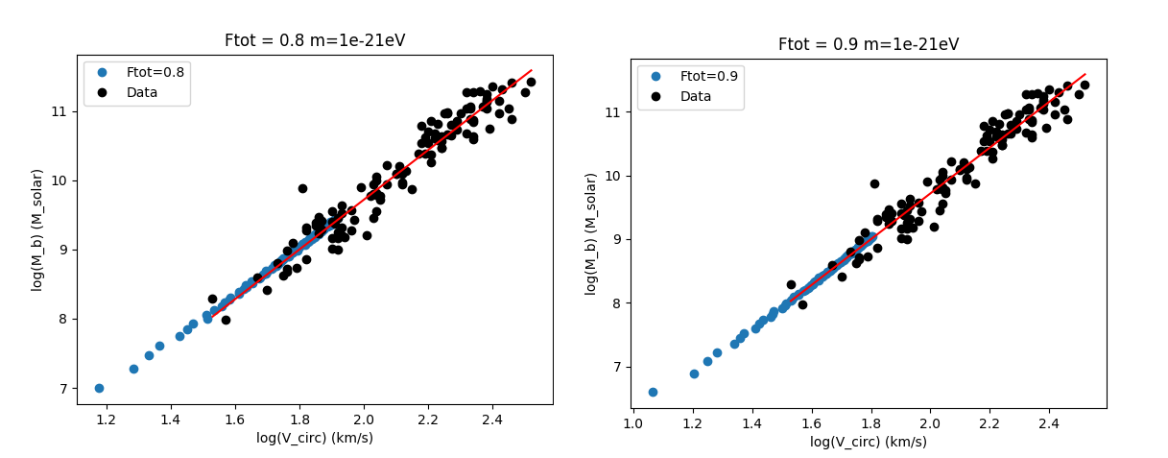}
\caption{A comparison of total dark matter fractions.  Increasing the total dark matter fraction shifts the simulated BTFR towards the higher mass regions; larger dark matter fractions produce large galaxies for fixed excitation number $n$.  Lower bounds on the mass parameter may be increased by considering higher values of the total DM fraction.}
\label{fig:FractionComparison}
\end{figure}

\begin{figure}
    \centering
    \includegraphics[height=3in,width=3.35in]{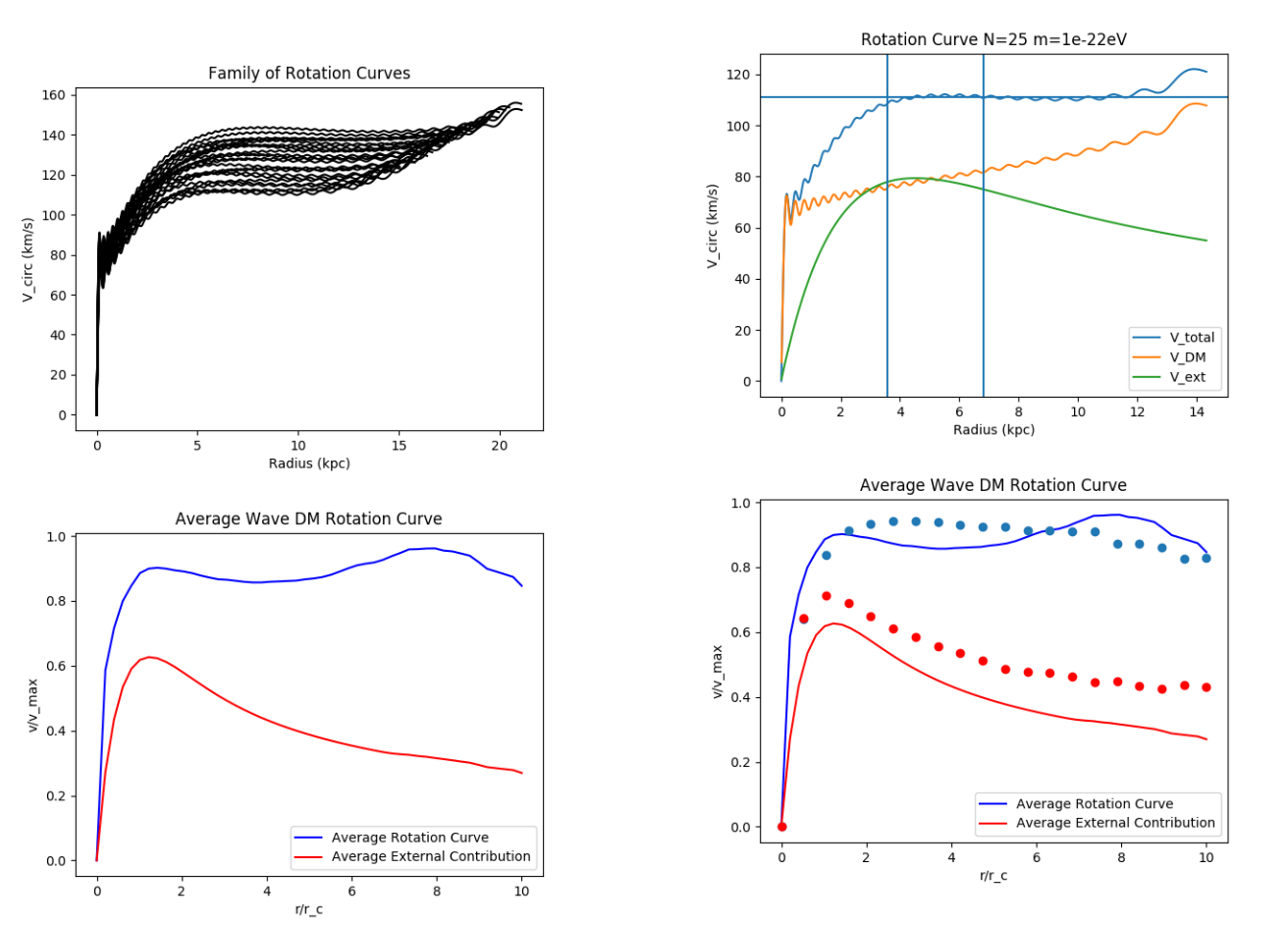}
    \caption{Rotation Curves:  Rotation Curves corresponding to the BTFR fit at $m=10^{-22}$eV, $f_{tot}=0.9$, and $f_{b} = 0.5$.  Units converted to kpc and km/s for realistic values.  Top left: A set of rotation curves from the family, orders $n=25-50$.  Top right: A single rotation curve along with its contributions from DM and the external component.  Vertical lines placed to denote $R_{b}$ and $\frac{R_d}{2}$, $R_{b}$ being the leftmost line. Horizontal line denotes the rotational velocity at the "flat part" of the curve, corresponding to the BTFR velocity.  Contributions computed as $v_{dm}(r) = \sqrt{\frac{M_{dm}(r)}{r}}$.  Bottom left: An averaged version of all $\psi$DM rotation curves corresponding to the fit.  Averages are computed similarly to those in figure \ref{fig:SPARCData}.  Bottom right: The same plot, but with an overlay of the averaged SPARC data.}
    \label{fig:RotationFamily}
\end{figure}

\begin{figure}
    \centering
    \includegraphics[height=3in,width=3.35in]{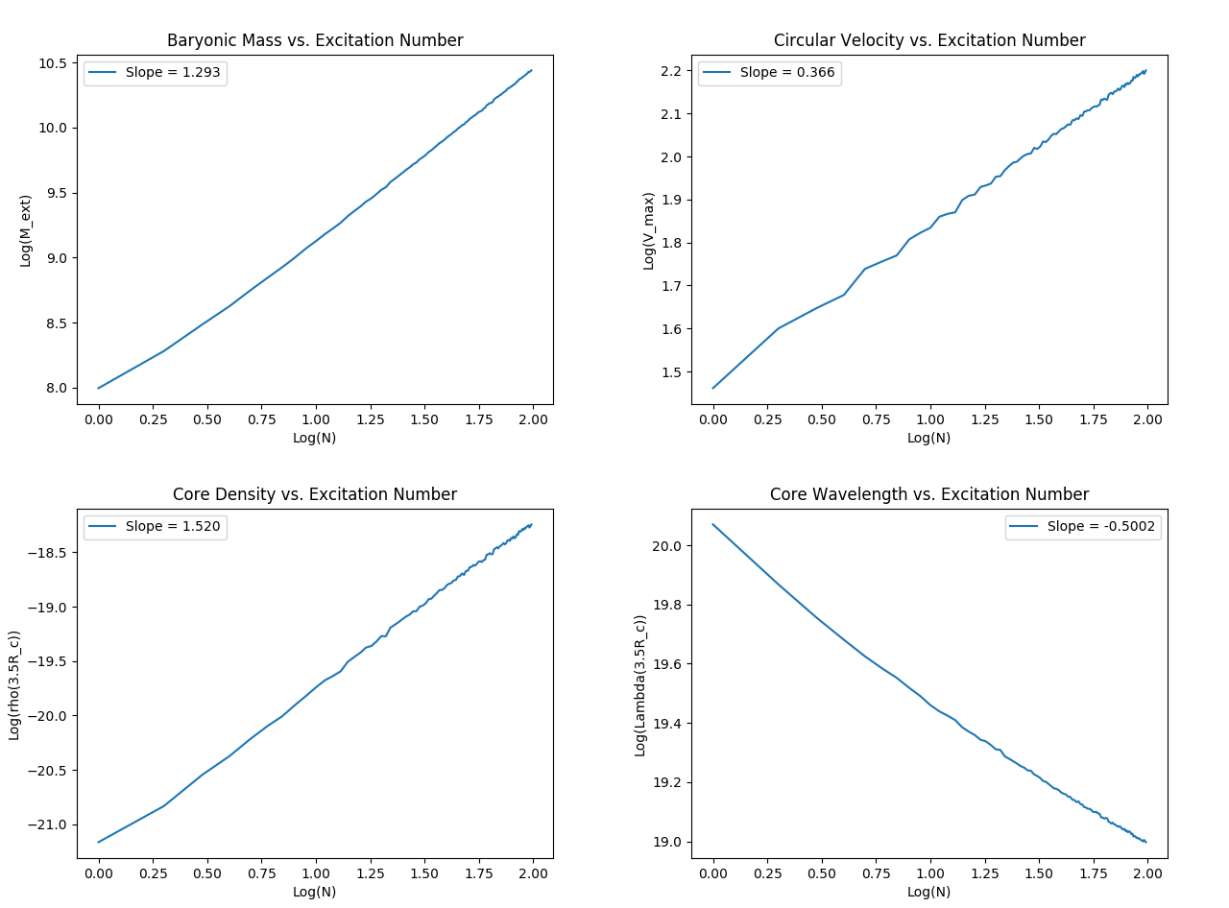}
    \caption{N dependence of galaxies for $m=10^{-22}$eV.  All axes are logarithmic. Top left:External mass vs. N.  Masses correspond to the BTFR fit in figure \ref{fig:TFComposite}.  Top right: Rotational velocity corresponding to the BTFR.  Bottom left: Density at the core boundary, $r=3.5r_c$.  Bottom right: Wavelength at the core boundary.  These relations allow one to estimate the excitation number of a halo given its mass or rotational velocity.  This has implications for the boundary of the central core, addressed in the discussion.}
    \label{fig:Ndependences}
\end{figure}

The results are summarized in figures (\ref{fig:TFComposite})-(\ref{fig:Ndependences}).  In figure (\ref{fig:TFComposite}) we display our best fitted results to the BTFR. Multiple values of particle mass, $m$, are displayed for fixed values of the defining dark matter fractions.   An example of varying the total dark matter fraction is displayed instead in figure (\ref{fig:FractionComparison}).  Varying the fraction, $f_b$ does not significantly affect the result; this feature is discussed in the following discussion section. 

An example of the resulting rotation curves appears in figure (\ref{fig:RotationFamily}).  In order to compare the rotation curves to the SPARC data, we perform the averaging procedure as described for the SPARC sample in figure (\ref{fig:SPARCData}) on the resulting curves.  The averaged $\psi$DM rotation curve is then compared to that of the SPARC sample.

We extract the dependence of various halo properties on the value of the state excitation number.  These relations are displayed on a logarithmic scale in figure (\ref{fig:Ndependences}).  These relations are then used to describe the mass and length scales of our DM Halos.  We extract a relation between the excitation number and the scale of density fluctuations at the boundary corresponding to the red boundary in figure (\ref{fig:BoundaryConditionNBody}).  We then use this scale to provide a comparison to high resolutions simulations and to provide a potential interpretation of our galactic boundary conditions.  In figure (\ref{fig:Ndependences}) we display the dependence of such fluctuations at the value of $r=3.5r_c$ where $r_c$ is defined as in equation (\ref{eq:CoreHalo}).  We will refer to this radius as the core boundary.  This choice is discussed in the following sections.

\section{Discussion}
\label{Sec:Discussion}
\subsection{Constraints on the $\psi$DM Particle Mass}
\label{Sec:MassConstraint}
An important feature of the BTFR fit is the location of each excited state on the BTFR plot.  Qualitatively, states of lower excitation number have lower masses and hence lower rotational velocities.  The lowest excited states thus provide a lower bound for the values of mass and velocity which are consistent with modelling galactic DM halos as SSS states.  This property is enhanced as the mass parameter is decreased.  In other words, decreasing $m$ leads to halos which are more and more massive for a given excitation number, $n$.  

We determine a bound of $m\geq 10^{-23}$eV.  This was determined by noting that some values of particle mass generate excited state galaxies of order $n\geq1$ which are too massive, and therefore insufficient in describing the lower mass end of the BTFR.  In other words, if $m$ is too small, the low mass region of the BTFR will not be covered by any of the excited states.  This constraint is consistent with most other published results, and thus reinforces the possibility that DM halos can be described to good approximation by SSS excited states.    

This constraint remains true even when varying the values of the dark matter fractions.  The value of the half-fraction (that at the radius from equation \ref{eq:Rhalf}) does not affect the overall fit to the BTFR.  This is due to the fact that the flat part of the rotation curve almost always occurs at distance much larger than the half mass radius of the external component, as seen in the top right of figure (\ref{fig:RotationFamily}).  This is consistent with the data from the SPARC survey.  As a result, the primary parameter which affects the fit is just the total fraction $f_{tot}$.  Adjusting $f_{tot}$ will affect the flat value of $v_{rot}$, and therefore slightly change the best fit to the BTFR; this is depicted in figure \ref{fig:FractionComparison}.  This dependence, given the small range of $f_{tot}$ considered in this analysis, is rather negligible, and does not significantly affect the lower bound on $m$.  Moreover, most galaxies are expected to have total fractions within this tested range.  We display two lower bound fits for values of $f_{tot}=.9$ and $f_{tot}=.8$ in figure \ref{fig:FractionComparison}.  

By strengthening our assumptions on the possible excitation numbers of BTFR galaxies, we can push our bound up to $m\geq10^{-22}$eV.  For instance, considering states of only $n\geq3$ pushes the bound up to the range of $m\geq10^{-22}eV$.  One could potentially use a similar idea to place an upper bound on the mass values compatible with the BTFR.  To achieve this, one may also claim an upper limit on the state excitation $n=N_{max}$, and consider a fixed range of excited states.  The simulated BTFR would then overlap data for a fixed range of particle mass.  Conversely, given a range of mass values, one may be able to place an estimate on the excitation numbers of Tully-Fisher type galaxies.  

We did not consider variations in the stellar light-to-mass ratio, $\Upsilon_*$, when computing the fit to the BTFR.  In reality, such variations could affect the fit dramatically, as doing so would cause a change in the slope and intercept of data itself.  However, we note that choosing the value of $\Upsilon_*=0.5$ as done in this analysis results in an excellent agreement between the data and the theoretically produced BTFR.  

\subsection{State Excitation Numbers}
\label{Sec:ExcitationNumbers}
An important feature of our model is the DM halo excitation number.  As described in the previous section, the excitation number of a given galaxy relates to its positioning on the BTFR, and therefore to its baryonic (external) mass content.  As seen in figure \ref{fig:Ndependences}, the baryonic mass relates to the state excitation number by an approximate power law
\begin{equation}
M_{ext} \underset{\sim}\propto N^{1.3}
\end{equation}
\begin{equation}
M_{ext}(N) \approx CN^{1.3}
\end{equation}
This dependence can be used to estimate the excitation numbers of a set of galaxies given their mass.  Further, one can determine how many excited states are required to cover the BTFR.  For instance, suppose we want to cover a mass range of 4 orders of magnitude (similar to the BTFR data used in this paper), starting with the first excited state, $N=1$.  From some simple algebra, one can determine the constant $C$ in the above equation to be $C=M_{ext}(1)$, and therefore recover a relation of
\begin{equation}
    \frac{M_{ext}(N)}{M_{ext}(1)} \approx N^{1.3}
\end{equation}
Then to cover the 4 orders of magnitude, we consider the value of $N$ required such that $\frac{M(N)}{M(1)} = 1000$.  This then implies that $N\approx 200$.  Thus, if we assume the lowest mass galaxy on the BTFR is in the state $N=1$, we require approximately 200 states to enclose the desired range of masses.  This is indeed reflected in figure \ref{fig:TFComposite}, showing that the BTFR is covered by the 200 computed states.  Equivalently one could use this estimate to determine the number of states required to span a range of rotational velocities as opposed to baryonic mass.

Claiming an upper limit to the excitation number would be a powerful tool in regards to placing constraints on the value of $m$.  In our analysis, we simply considered the lower bound on $N$, resulting in a lower bound on $m$.  It is also possible to consider placing an upper bound on $N$; this would result in an upper bound on $m$ to ensure compatibility with the BTFR.  Combining this with the lower bound, one would then find that for each pair of $N_{min}$ and $N_{max}$ there is a range of $m$ values which allows the simulated halos to overlap with the BTFR.  We did not place an upper bound on $N$, as there is not yet a clear mechanism to explain this; any mechanism for placing an upper bound on $N$ at this point is purely speculative.

\subsection{Modelling of Individual Rotation Curves}
\label{Sec:IndividualCurves}
The fact that appropriately chosen families of SSS states are able to reproduce the BTFR suggests that SSS states could be used to model the rotation curves of individual galaxies.  We display a comparison of our simulated curves to those from SPARC in figure \ref{fig:RotationFamily}.  Similar to the SPARC data, curves with $m\approx10^{-22}$eV display a rise on the order of 1kpc, and extend to the order of 10kpc.  The baryonic component is almost completely enclosed before the maximal velocity is achieved.  We display the average $\psi$DM rotation curve with the averaged SPARC curve as a qualitative comparison.  While the baryonic components do not display a perfect correspondence, this could be improved with a more rigorous treatment of both the stellar-light-to-mass ratio $\Upsilon_*$ and the DM fraction at the stellar half mass radius $f_{b}$.  We note that values of $f_b\approx 0.5$ most closely resemble the features of actual rotation curves.  That is, the DM and baryon contributions must be of similar scale at the stellar half mass radius, $R_b$, in order to display appropriate flattening of the rotation curve.

In this paper, we use a simple exponential model which corresponds to the baryonic component of the galaxy; while this is a theoretically convenient choice, it choice greatly affects the resulting shape of the rotation curve.  The rotation curves from this paper then would only be viable candidates for galaxies whose baryonic components are approximated well by this exponential model.  We note that this density model, equation \ref{eq:rhoext}, is a \textit{spherically symmetric approximation}.  In reality, galaxies do not display this strict spherical symmetry.  This is likely one source of the discrepancy between our average $\psi$DM rotation curve, and the average from SPARC displayed in figure \ref{fig:RotationFamily}. We note that given an improved model of the external component, the methods described in section \ref{Sec:SSSGalaxies} remain valid if and only if one strictly retains this requirement of spherical symmetry.  

Importantly, we do not expect the shortcomings of the spherically symmetric exponential model to significantly affect the resulting fit to the BTFR.  This is due to the DM halos having a much greater spatial extent and mass fraction than the external component.  In other words, the external component is always completely enclosed before the radius at which the maximal circular velocity is achieved.  As a result, the maximal rotational rate extracted for the BTFR is marginally affected by the shape and distribution of the external component.  This is indeed representative of the SPARC sample.  In other words, two equivalently massed external components will generate similar velocities in the outer regions where the curve is flat, reproducing the same data point on the BTFR.  Improving the model of the external component will affect the central regions of the rotation curve greatly, while minimally changing the features of the outer regions relevant to the BTFR. 

The length scale of the halos generated depends on the mass parameter $m$.  Loosely speaking, lower values of mass correspond to galaxies with greater spatial extent. This can be understood by considering the scaling relations for $m$ displayed in section \ref{Sec:ScalingRelations}. Rescaling to a lower $m$ value corresponds to choosing a value of $\beta<1$; this then results larger values of $r$ after rescaling. Importantly, $m$ is not the only important variable when it comes to this length scale.  The value of the DM fraction can also have significant effect.  In general, larger values of $f_{tot}$ will correspond to galaxies with larger spatial extent.  That is, converting some of the baryonic mass into DM mass will increase the outwards scalar pressure of the halo and cause it to expand.  Conversely, converting some of the DM mass into baryonic mass results in a contraction of the halo.

\subsection{Interpretation of Boundary Conditions}
\label{Sec:BoundaryInterpretation}
In section \ref{Sec:AmpWaveengthBoundary} we discussed applying a generic boundary condition, namely one which sets a common wavelength and amplitude scale amongst our $\psi$DM excited states.  We noted that this boundary problem relied on constructing an invariant quantity $I_{DM}=\lambda_{DM}^2A_{DM}$ for each excited state.  Each state was scaled to have the same value of $A_{DM}$ at the radius which achieves the value of $I_{DM}$ for that state.  In a sense, these ''boundary" radii are a computational convenience; they pick out the locations of the states which can be similarly scaled in both amplitude and wavelength.  The physical interpretation of these ''boundary" radii is less clear and requires additional investigation.  In fact, most of these radii occur near the decay radius, $R_d$ which occurs in the outer regions of the halo.  In comparison to the figure \ref{fig:BoundaryConditionNBody} which inspired our discussion of boundary conditions, we would like to instead consider some boundary closer to the core region.  In the following, we present a potential interpretation of this region based on the observations of our simulations.    

Returning to a more qualitative discussion, we suggest a possible interpretation of the amplitude-wavelength boundary condition.  Returning to figure \ref{fig:BoundaryConditionNBody}, we interpret the outer particle-like region of the halo as a fluctuating turbulent structure.  We suggest using the amplitude-wavelength condition to place an estimate on the size and density scale of the turbulent fluctuations.  If one identifies a boundary between the inner core of the halo and the outer turbulent region, then one could further identify the amplitude and wavelength at this boundary with those of the fluctuations. 

Following observations from the simulations of\cite{MoczGalaxy} that such a break occurs near $r=3.5r_{c}$, we consider this value of the boundary radius.  Again, $r_c$ is defined as in equation \ref{eq:CoreHalo}, to be the radius at which the DM density reaches half of it's central value.  We offer no analytical reason to choose the multiple of $3.5r_{c}$, but merely wish to make a comparison to high detailed simulations.  The dependence of the wavelength and density at this value of $r$ is displayed in figure \ref{fig:Ndependences}.  Interestingly we recover relationships which are close to rational, 
\begin{equation}
    \label{eq:proportion1}
    \lambda(3.5r_c) \underset{\sim}\propto N^{-0.5}
\end{equation}
\begin{equation}
\label{eq:proportion2}
\rho(3.5r_c) \underset{\sim}\propto N^{1.5}
\end{equation}
Combining these relations demonstrates another interesting relation, namely
\begin{equation}
    \rho(3.5r_c)\underset{\sim}\propto \lambda^{-3}
\end{equation}
This relation offers a possible interpretation for the boundary conditions which generate the TFR.  In particular, we can describe the mass scale of fluctuations near this boundary, $M_{fluctuation} \approx \frac{4\pi}{3}\rho\left(\frac{\lambda}{2}\right)^3$.  The above relations imply that this mass scale is held constant amongst the simulated halos. In other words, holding the mass scale of fluctuations constant near the boundary of the core region reproduces Tully-Fisher scalings.  
In the case of $m=10^{-22}$eV, we extract this particular mass scale and find that $M_{fluctuation}\approx 10^{8}-10^{9}M_{\odot}$. We now return to the possibility that these fluctuations may condense into $\psi$DM solitons at sufficient distance from the core.  If this were to occur, then the corresponding soliton size determined from the relation \ref{eq:GSMass} is of the order $R_{1/2} \approx 0.3 kpc$, which is comparable to the length scale of globular clusters.    We suggest that this mass scale could be related to the that of the halo's parent solitons, being approximately preserved during the formation of the halo.  Moreover, in practice one may wish to identify the radii such that the proportions \ref{eq:proportion1} and \ref{eq:proportion2} are exact.  This could provide a more rigorous definition of the boundary between the inner and outer regions of the halo.

\section{CONCLUSIONS}
\label{Sec:Conclusion}
$\psi$DM depicts galactic DM halos as an intricate and turbulent wave structure.  This idea allows one to consider the shape and behavior of $\psi$DM halos.  We considered the possibility of using SSS excited SF$\psi$DM states as a leading order model for DM halos.  By considering the inclusion of appropriate fractions of external matter to the SSS states as well as the appropriate boundary conditions, we demonstrate compatibility with the BTFR in the mass range $m\geq10^{-22}$eV.  The imposed boundary conditions imply a characteristic mass scale which is fixed amongst the SSS SF$\psi$DM excited states.  This model provides a new theoretical mechanism for producing the BTFR which has not been displayed in other DM models; this is particularly interesting, and may have implications in regards to the formation and morphology of galaxies.  It may also be possible to utilize the BTFR to estimate the order of a SF$\psi$DM halo's excitation number.  Our analysis also indicates SSS excited states to be viable candidates for producing rotation curve models for individual galaxies; this idea could be further investigated through an extension of the model in this paper.

\bibliographystyle{ieeetr}
\bibliography{Bibliography1.bib}
\section{Appendix: Numerical Solving of the SSS Excited States}

Here we focus on computing numerical solutions to the SSS EKGEs.  We present this computation similarly to the thesis of Andrew Goetz \cite{GoetzThesis}.  As a starting point, we list the SSS PSEs as in equations \ref{eq:SSSSchrodinger}-\ref{eq:SSSMass}.

To generate physically reasonable solutions, we must take the following set of assumptions:
\begin{equation}
\label{Eq:Criterion1}
  M(0)=\Phi_r(0) = 0
\end{equation}
\begin{equation}
\label{Eq:Criterion2}
 \lim_{r\to\infty}M(r) = M_{\infty}<\infty
\end{equation}
\begin{equation}
\label{Eq:Criterion3}
 \lim_{r\to\infty}V(r) = 0
\end{equation}

The first assumption is necessary to ensure regularity at the origin, $r=0$.  This can be concluded from considering the $\frac{M}{r^2}$ and $\frac{2\Phi_r}{r}$ terms in equations 2 and 3. Next, the assumption on $M(r)$ ensures that solutions have finite mass contained in the dark matter scalar field.  Lastly, the assumption on $V(r)$ corresponds to taking the convention that the gravitational potential approaches $0$ at infinite distance from the origin.  It should be noted that an arbitrary constant, $\bar{V}$, can be added to $V(r)$ without affecting the solution.  Thus, a solution which satisfies the first two assumptions, but not the third can be made to satisfy the third by an appropriate adjustment of $V(r)$.    

Each solution is then specified by a choice of initial conditions, $(\Phi_0,V_0) = (\Phi(0),V(0))$, as well as its frequency $\omega$.  We take the convention that the potential function, $V(r)$, is always negative, and thus consider only $V_0<0$.  Moreover, it can be seen that given a solution for $(\Phi(r),M(r),V(r))$, that $(-\Phi(r),M(r),V(r))$ also generates a solution.  For this reason we choose to always take the value of $\Phi_0$ to be positive without any loss of generality. 

Numerically, it is convenient to consider the case in which the frequency $\omega<m$ is taken to be fixed.  Solutions with fixed $\omega$ can then be uniquely specified by their excitation number $n$.  Finding a solution of order $n$ then requires one find the appropriate initial conditions, which will depend on $n$ and $\omega$, $(\Phi_0(n,\omega),V_0(n,\omega)$.  These initial conditions will then produce solutions $(\Phi(r;n,\omega),V(r;n,\omega),M(r;n,\omega))$ which must obey the criteria \ref{Eq:Criterion1}-\ref{Eq:Criterion3}. 

Taking a naive guess of $(\Phi_0,V_0)$ will likely result in a solution which diverges exponentially at large radii, violating the second and third requirements, \ref{Eq:Criterion2} and \ref{Eq:Criterion3}.  In fact, for a fixed value of $\Phi_0$, there are a countable number of values for $V_0$ which do not diverge for large radii; these $V_0$ correspond to bound, finite mass, excited states, but will not satisfy the convention for $V_\infty$ in general.  To make sure the condition for $V_{\infty}$ is satisfied, one may consider the asymptotic behavior of $V(r)$.  That is, for $V$ to appropriately approach $0$ in the Newtonian fashion, it must satisfy the following condition: 
\begin{equation}
V(r) - \frac{1}{2} \ln\left(1-\frac{2M(r)}{r}\right) = Y(r) \approx 0
\end{equation}
This condition is equivalent to the assumption that the spacetime metric is asymptotically Schwarszchild, or in other words, that the potential $V(r) \approx \frac{M(r)}{r}$ at large radii.   

Finding the correct set of initial conditions can be achieved through a shooting problem-like method.  Such a method, further detailed in \cite{GoetzThesis}, is achieved as follows: 
\begin{itemize}
    \item Choose a value of $\omega<m$, and guess a value for $\Phi_0$.  
    \item Given $\Phi_0$, choose a value of $V_0$ which is consistent with the condition of $k^2(0)<0$, this ensures the wavefunction is initially oscillatory as are the expected solutions.  
    \item Solve the ODES \ref{eq:SSSSchrodinger}-\ref{eq:SSSMass} up to the decay radius $R_d$ where $k(R_d) = 0$, and extend the solution to a chosen tolerance past $R_d$.  Count the number of zeros $\Phi$ displays up to this point, and denote it as $N$.
    \item To generate a solution of order $n$, adjust the value of $\Phi_0$ until $N=n$, record this value of $\Phi_0$ as $\Phi_n$.
    \item Further adjust $\Phi_0$ to attain a solution with $N=n+1$, and record $\Phi_{n+1}$
    \item The pairings $(\Phi_n,V_0)$ and $(\Phi_{n+1},V_0)$ generate solutions with $n$ and $n+1$ zeros, though may display exponential divergences.
    \item To find the set without exponential divergence, perform a bisection search in the value of $\Phi_0$.  The bound, non-diverging solution with $n$ zeros will lie on the boundary of solutions with $n$ and $n+1$ zeros. Call the resulting value of $\Phi_0=\Phi_1$.
    \item The pair $(\Phi_1,V_0)$ generates a bound solution of order $n$, but may not have $V_{\infty}=0$
    \item To achieve $V_{\infty}=0$, perform a shooting problem in the value of $V_0$, repeating the entirety of the above procedure for each considered value of $V_0$. Note that the previous value of $\Phi_1$ will provide a good initial guess for the next iteration.  
    \item Vary $V_0$ and repeat procedure until the condition \ref{Eq:Criterion3} is satisfied to a determined tolerance.  
    \item Result is $(\Phi_0(n,\omega),V_0(n,\omega))$.  To generate solutions for different values of $\omega$ (and therefore of different mass scale), apply the scaling relations from section \ref{Sec:ScalingRelations}.  

\end{itemize}

This procedure as outlined, will generate SSS solutions to the DM-only PSEs.  However, solutions which include external potentials due to other matter are quite analogous.  The necessary conditions in \ref{Eq:Criterion1}-\ref{Eq:Criterion3} can still be achieved in this setting with the procedure outlined above, though the use of a continuation parameter as stated in \ref{Sec:ModelDetails} greatly simplifies the problem.  To include an external potential then, one may repeat the above procedure first in a DM-only setting, and then slowly introduce the external potential.  That is, considering $V_{tot} = V+\alpha V_{ext}$, the solution is close to that of the DM-only setting if $\alpha$ is taken sufficiently small.  Therefore, utilizing small steps in $\alpha$ one may iterate the outlined procedure until $\alpha=1$, fully including the external contribution.

\end{document}